\documentclass[10pt,twocolumn,aps,prd,preprintnumbers,showpacs,superscriptaddress,nofootinbib,amsmath,amssymb,floats,floatfix,showkeys,notitlepage,longbibliography]{revtex4-2}
\usepackage{orcidlink}
\usepackage{comment}
\usepackage{lipsum}
\usepackage{graphicx}
\usepackage{subfigure}
\usepackage{palatino}
\usepackage{sans}
\usepackage{hyperref}
\hypersetup{colorlinks=true,linkcolor=blue,urlcolor=blue,citecolor=blue}
\usepackage[toc,page]{appendix}
\usepackage[normalem]{ulem}
\usepackage{adjustbox}
\usepackage{latexsym}
\usepackage{amsmath}
\numberwithin{equation}{section}
\usepackage{amssymb}
\usepackage{amsfonts}
\usepackage{dcolumn}
\usepackage{bm}
\usepackage{tikz}
\usetikzlibrary{decorations.pathmorphing}
\usepackage{bigints}
\usepackage{array,tabularx,multirow,booktabs}
\usepackage[tracking=true]{microtype}
\usepackage{soul} 
\SetTracking{}{500}
\SetTracking{encoding={*}, shape=sc}{40}
\UseRawInputEncoding 
\allowdisplaybreaks
\usepackage[utf8]{inputenc}
\usepackage{xcolor} 
\newcommand{\nn}{\nonumber \\}

\begin{document} \sloppy

\title{Worldline-jet affine spectral approach for near-horizon atomic radiation and HBAR entropy}

\author{Reggie C. Pantig \orcidlink{0000-0002-3101-8591}}
\email{rcpantig@mapua.edu.ph}
\affiliation{Physics Department, School of Foundational Studies and Education, Map\'ua University, 658 Muralla St., Intramuros, Manila 1002, Philippines.}

\author{Uktamjon Uktamov} 
\email[Corresponding author:]{uktam.uktamov11@gmail.com}
\affiliation{School of Physics, Harbin Institute of Technology, Harbin 150001, People’s Republic of China}
\affiliation{Institute for Advanced Studies, New Uzbekistan University, Movarounnahr str. 1, Tashkent 100000, Uzbekistan}

\author{Ali \"Ovg\"un \orcidlink{0000-0002-9889-342X}}
\email{ali.ovgun@emu.edu.tr}
\affiliation{Physics Department, Eastern Mediterranean University, Famagusta, 99628 North
Cyprus via Mersin 10, Turkiye.}

\author{Bobomurat Ahmedov \orcidlink{0000-0002-1232-610X}}
\email{ahmedov@astrin.uz}
\affiliation{School of Physics, Harbin Institute of Technology, Harbin 150001, People’s Republic of China}
\affiliation{Institute of Theoretical Physics, National University of Uzbekistan, Tashkent 100174, Uzbekistan.}

\begin{abstract}
In this paper, we develop a local worldline-jet spectral treatment of atomic radiation near static black-hole horizons. For Schwarzschild radial infall, a logarithmic outgoing phase yields a Fourier-Mellin integral and closed gamma-function emission and absorption amplitudes. Their high-gap ratio recovers established HBAR thermal detailed balance. Higher worldline moments give explicit finite-interaction corrections and nonthermal spectral curvature. A photon-number master equation yields a geometric stationary distribution and an exact entropy balance, with the familiar HBAR area coefficient recovered under conditional radiative assumptions. We then apply the same geodesic-jet and Mellin-moment procedure to the nondegenerate Simpson-Visser black bounce. The first two regular phase coefficients depend on the bounce parameter and infall energy. At a distinguished nondegenerate parameter value, the linear coefficient vanishes while the quadratic coefficient retains an energy-squared dependence. At the critical null throat, the logarithmic phase is replaced by an inverse-proper-time phase and the local spectral integral takes a modified-Bessel form. The calculations concern idealized outgoing modes and do not establish a complete global radiation spectrum.
\end{abstract}


\maketitle

\section{Introduction}\label{sec1}

Quantum fields in curved spacetime admit particle interpretations that depend on the quantum state and the time evolution used to define positive frequency. The Schwarzschild geometry provides a particularly direct setting for this distinction through the Hawking radiation process, the exterior Boulware vacuum, and the response of accelerated or freely falling observers \cite{Fulling:1972md,Davies:1974th,Boulware:1974dm,Hawking:1975vcx,Unruh:1976db,Crispino:2007eb}. The associated surface-gravity temperature is central to black-hole thermodynamics, while the entropy assigned to a radiation subsystem must be distinguished from the intrinsic entropy of the horizon \cite{Bekenstein:1973ur,Page:1976df}. An atom with two internal energy levels provides an operational way to probe the field state along a specified worldline without requiring a global particle interpretation.

Quantum-optical models of accelerated atoms have long emphasized the counter-rotating process in which atomic excitation accompanies photon creation. Cavity boundaries and switching protocols can influence this process, even when the conventional Unruh factor appears in an asymptotic regime \cite{Scully:2003zz,Hu:2004kg,Belyanin:2004cvu}. Building on the quantum theory of laser-like systems \cite{PhysRevLett.16.853}, Scully and collaborators studied atoms freely falling into a Schwarzschild black hole through a Boulware-like exterior state. The resulting horizon-brightened acceleration radiation, or HBAR, was distinguished from ordinary Hawking radiation, and a near-steady photon entropy flux was related to an area term \cite{Scully:2017utk}. The comparison with an accelerated mirror likewise illustrates the relevance of counter-rotating atomic transitions and the difference between atomic and field-mode frequencies \cite{Svidzinsky:2018jkp}.

Near-horizon conformal quantum mechanics has already supplied an analytical account of HBAR thermality and its relation to the surface gravity \cite{Camblong:2004ye,Camblong:2020pme}. Extensions to randomly injected atomic clouds and rotating horizons derive thermal photon occupations and HBAR thermodynamic relations from multimode master equations \cite{Azizi:2020gff,Azizi:2021qcu,Azizi:2021yto}. Further investigations include quantum-corrected geometries, dark-energy backgrounds, generalized-uncertainty modifications, regular black holes, and massive vector fields \cite{Sen:2022tru,Bukhari:2022wyx,Ovgun:2025isv,Ovgun:2025ehi,Pantig:2025eew}. The Simpson-Visser black-bounce family interpolates between a nondegenerate black hole and a critical null throat \cite{Simpson:2018tsi}. The thermal nonextremal HBAR response and its entropy have also been studied in that family \cite{Ovgun:2026plj}. These studies establish that obtaining a Planck factor or an area coefficient is not, by itself, a distinguishable new physical result.

Detector spectra need not remain thermal when the interaction is limited in time or when nonstationary details are retained. Finite detector extent, smooth switching, and the order of long-time limits have been studied systematically in the Unruh-DeWitt framework \cite{Schlicht:2003iy,Louko:2006zv,Satz:2006kb,Louko:2007mu}. Other trajectory and Schwarzschild detector studies distinguish local response functions from global quantum states \cite{Barbado:2012fy,Hodgkinson:2014iua}. In the more immediate HBAR setting, Lobos and Rodulfo have obtained finite-gate transition expressions for a Dirac detector in a Schwarzschild-MOG background while retaining both near-horizon scattering branches and their interference \cite{Lobos:2026ahg}. Their analysis makes particularly clear why a local outgoing wave cannot be identified with an entire globally normalized mode.

A complementary mathematical language arises from the logarithmic relation between outgoing Schwarzschild retarded time and the proper time remaining before a radial geodesic crosses the horizon. Translating the retarded coordinate produces an asymptotic dilation of the remaining proper time, so that a Fourier phase is represented by an imaginary power suited to Mellin analysis. Affine-group formulations of the Unruh effect and related modular analyses have already made the translation-dilation link explicit \cite{Arzano:2025aeo,Arzano:2026vwz}. Our objective is not to rediscover that group structure, but to examine what follows when successive derivatives of the actual Schwarzschild infall trajectory are retained within a single local spectral calculation.

We develop a complementary local spectral treatment of the HBAR transition problem studied by Scully and collaborators. Rather than claim the established Planck factor or leading entropy-area coefficient as new, we use the proper-time jet of the infalling trajectory to organize emission and absorption within a common family of Mellin-Laplace moments. We examine how finite interaction width and nonlinear worldline terms enter the transition quotient, and how that quotient appears in a photon-number entropy identity under specified radiative assumptions. We also apply the same procedure to the nondegenerate and critical regimes of the Simpson-Visser geometry. The resulting probabilities refer to a selected outgoing mode and are not identified with a complete cavity response.

We first establish the radial worldline expansion and its retarded-time scaling, then derive the affine spectral transition amplitudes. The thermal high-gap regime is examined before successive geometric corrections and spectral curvature are considered. A birth-death description subsequently connects the transition quotient to photon statistics, entropy balance, and a radiative area equivalent. We then apply the same spectral moments to the nondegenerate Simpson-Visser horizon and examine how the critical null throat changes the phase integral. The discussion closes by identifying the domains in which these results may be interpreted. Throughout, we use the metric signature \((-+++)\), geometric units \(G=c=\hbar=k_B=1\) unless otherwise indicated, Schwarzschild mass \(M>0\), surface gravity \(\kappa=1/(4M)\), a fixed positive infall energy per unit rest mass, and positive atomic and outgoing Killing frequencies. The selected field is a massless scalar in an exterior Boulware-like outgoing-mode approximation.

\section{Schwarzschild Geometry and the Near-Horizon Worldline-Jet Expansion}
\label{sec2}

We consider radial infall into a nonrotating black hole of mass \(M>0\) and work in geometric units \(G=c=\hbar=k_B=1\). The Schwarzschild radius and surface gravity are \(r_s=2M\) and \(\kappa=1/(2r_s)\). In the exterior region \(r>r_s\), the line element takes the form
\begin{align} d\ell^2&=-f(r)\,dt^2+\frac{dr^2}{f(r)}+r^2d\Omega_2^2,\nn f(r)&=1-\frac{r_s}{r}. \label{2.1} \end{align}
The angular metric is \(d\Omega_2^2=d\theta^2+\sin^2\theta\,d\phi^2\). The Schwarzschild time \(t\) refers to the static Killing field, whereas \(\tau\) denotes the proper time of a freely falling atom. The radial trajectories considered here extend smoothly through the future horizon in regular coordinates. Their exterior portions suffice for the outgoing-wave phase relevant to the near-horizon atomic response studied in HBAR \cite{Boulware:1974dm,Scully:2017utk,Camblong:2020pme}.

\subsection{Radial infall in the remaining proper time}

For a future-directed timelike radial geodesic, the conserved specific Killing energy \(E>0\) and timelike normalization give
\begin{align} E&=f(r)\frac{dt}{d\tau},\nn \left(\frac{dr}{d\tau}\right)^2&=E^2-f(r). \label{2.2} \end{align}
We choose the infalling branch \(dr/d\tau=-\sqrt{E^2-f(r)}\). Let \(\tau_H\) be the finite proper time of horizon crossing and introduce the positive remaining proper time \(s=\tau_H-\tau\). Thus \(s\to0^+\) means approach to the horizon from outside, and the radial equation becomes
\begin{align} \frac{dr}{ds}&=\sqrt{E^2-f(r)},\nn r(0)&=r_s,\nn \frac{dr}{ds}\bigg|_{s=0}&=E. \label{2.3} \end{align}
The square root is taken on its positive branch. For every fixed \(E>0\), its argument remains positive in a sufficiently small exterior neighborhood of the horizon. The differential equation therefore admits an ordinary Taylor expansion about \(s=0\), even though Schwarzschild time is singular there.

Differentiating the first relation in Eq. \eqref{2.3} eliminates the square root from the second derivative. A further derivative then gives
\begin{align} \frac{d^2r}{ds^2}&=-\frac{1}{2}f'(r)=-\frac{r_s}{2r^2},\nn \frac{d^3r}{ds^3}&=\frac{r_s}{r^3}\frac{dr}{ds}. \label{2.4} \end{align}
At horizon crossing, these identities imply
\begin{align} r(0)&=r_s,\nn r'(0)&=E,\nn r''(0)&=-\frac{1}{2r_s},\nn r'''(0)&=\frac{E}{r_s^2}. \label{2.5} \end{align}
The resulting worldline jet is
\begin{equation} \frac{r(s)}{r_s}=1+E\frac{s}{r_s}-\frac{1}{4}\left(\frac{s}{r_s}\right)^2+\frac{E}{6}\left(\frac{s}{r_s}\right)^3+\mathcal{O}_E\!\left(\frac{s^4}{r_s^4}\right). \label{2.6} \end{equation}
The subscript on \(\mathcal{O}_E\) indicates that the remainder constant may depend on the fixed geodesic energy. The linear coefficient records the horizon-crossing radial velocity in the chosen proper-time variable. The quadratic coefficient follows from the radial geodesic acceleration and is independent of \(E\). No assumption that the atom was released from rest at infinity has entered the expansion.

The case \(E=1\), corresponding to radial fall from rest at infinity, provides an independent analytic check. Direct integration of Eq. \eqref{2.3} yields
\begin{equation} \frac{r(s)}{r_s}=\left(1+\frac{3s}{2r_s}\right)^{2/3}. \label{2.7} \end{equation}
Its binomial expansion reproduces every displayed coefficient of Eq. \eqref{2.6} at \(E=1\). The local series for general \(E\) is therefore consistent with an exactly integrable reference trajectory without relying on that special choice in the subsequent analysis.

\subsection{Outgoing retarded time and its logarithmic jet}

Introduce the Schwarzschild tortoise radius \(r_*\) and the outgoing and ingoing null coordinates \(u\) and \(v\) through
\begin{align} r_*&=r+r_s\ln\!\left(\frac{r-r_s}{r_s}\right),\nn u&=t-r_*,\nn v&=t+r_*. \label{2.8} \end{align}
The logarithm has a positive, dimensionless argument throughout the exterior. Since \(dr_*/dr=f(r)^{-1}\) and \(dt/ds=-E/f(r)\), differentiation along the infalling worldline gives
\begin{align} \frac{du}{ds}&=-\frac{E+\sqrt{E^2-f(r(s))}}{f(r(s))},\nn \frac{dv}{ds}&=-\frac{E-\sqrt{E^2-f(r(s))}}{f(r(s))}. \label{2.9} \end{align}
Rationalizing with \(f=E^2-(E^2-f)\) produces the equivalent expressions
\begin{align} \frac{du}{ds}&=-\frac{1}{E-\sqrt{E^2-f(r(s))}},\nn \frac{dv}{ds}&=-\frac{1}{E+\sqrt{E^2-f(r(s))}}. \label{2.10} \end{align}
The different horizon limits are now explicit. The outgoing derivative diverges as \(s\to0^+\), whereas the advanced coordinate remains regular and satisfies \(dv/ds\to-1/(2E)\). The logarithmic behavior of \(u\) is a property of the outgoing Schwarzschild null phase, not a divergence in the proper time or the atom's four-acceleration.

For a systematic expansion, set \(y=s/r_s\) and substitute Eq. \eqref{2.6} into \(f=1-r_s/r\). The result is
\begin{equation} f(r(s))=Ey-\left(E^2+\frac{1}{4}\right)y^2+\left(E^3+\frac{2E}{3}\right)y^3+\mathcal{O}_E(y^4). \label{2.11} \end{equation}
Using this series together with \(dr/ds\) from Eq. \eqref{2.3} in the first identity of Eq. \eqref{2.9}, we obtain
\begin{equation} \frac{du}{ds}=-\frac{2r_s}{s}-2E+\frac{s}{3r_s}+\mathcal{O}_E\!\left(\frac{s^2}{r_s^2}\right). \label{2.12} \end{equation}
The cancellation of every \(E\)-dependent contribution to the coefficient of \(s\) is a consequence of the geodesic equations and the exact tortoise-coordinate relation. Integrating term by term over a sufficiently small positive interval gives
\begin{equation} u(s)=u_0-\frac{1}{\kappa}\ln\!\left(\frac{s}{s_0}\right)-2Es+\frac{s^2}{6r_s}+\mathcal{O}_E\!\left(\frac{s^3}{r_s^2}\right). \label{2.13} \end{equation}
Here \(s_0>0\) is a reference proper-time scale and \(u_0\) is an additive coordinate reference. Only their combined choice fixes the origin of \(u\), so changing \(s_0\) can be compensated by shifting \(u_0\). Both the coefficient \(1/\kappa=2r_s\) of the logarithm and the quadratic coefficient \(1/(6r_s)\) are independent of the radial geodesic energy. By contrast, the first regular correction, \(-2Es\), retains information about the atom's trajectory.

The incoming coordinate provides a regular counterpart. Direct integration of the second identity in Eq. \eqref{2.10} gives a finite horizon value \(v_H\) and yields
\begin{equation} v(s)=v_H-\frac{s}{2E}+\mathcal{O}_E\!\left(\frac{s^2}{r_s}\right). \label{2.14} \end{equation}
Thus the outgoing and ingoing wave phases have qualitatively different proper-time behavior along the same freely falling trajectory. The singular logarithm in Eq. \eqref{2.13} does not indicate a singular infalling worldline.

\subsection{Affine scaling of the outgoing phase}

In the near-horizon geometric-optics sector, an exterior outgoing positive-frequency mode has local retarded-time dependence \(e^{-i\nu u}\) with Killing frequency \(\nu>0\). Define the dimensionless ratio \(\sigma=\nu/\kappa\). Along the trajectory, Eq. \eqref{2.13} implies
\begin{align} e^{-i\nu u(s)}&=e^{-i\nu u_0}\left(\frac{s}{s_0}\right)^{i\sigma}\nn &\exp\!\left(2iE\nu s-\frac{i\nu s^2}{6r_s}+\mathcal{O}_E\!\left(\frac{\nu s^3}{r_s^2}\right)\right). \label{2.15} \end{align}
The logarithmic term turns the outgoing phase into an imaginary power of the remaining proper time. At leading order, a fixed translation of \(u\) dilates \(s\). The exterior map \(s\mapsto u(s)\) is monotone and has a local inverse \(s(u)\). For fixed \(\Delta u\), we obtain
\begin{equation} \lim_{s\to0^+}\frac{s(u(s)+\Delta u)}{s}=e^{-\kappa\Delta u}. \label{2.16} \end{equation}
The translation-dilation correspondence is the geometric basis for the affine spectral description. A Fourier phase in \(u\) becomes a Mellin-type power in \(s\), while successive worldline-jet coefficients supply corrections with explicit powers of the remaining proper time. The relation is asymptotic rather than an assertion of exact dilation symmetry at finite distance from the horizon.

The expansions assume a fixed \(E>0\) and \(0<s/r_s\ll1\), with \(f(r(s))\ll E^2\) ensuring a regular expansion of the radial square root. The constants implicit in the remainders need not be uniform as \(E\to0^+\). For a high-frequency wave, truncation of the phase in Eq. \eqref{2.15} additionally requires the discarded phase contribution \(\nu s^3/r_s^2\) to remain small in the region relevant to the transition amplitude. No choice of global vacuum, radial scattering modes, or cavity boundary conditions follows from this local geometric calculation. Those physical inputs must be specified independently when the outgoing phase is used in an atom-field transition integral.

\section{Affine Spectral Representation of Atomic Transitions}
\label{sec3}

We describe a freely falling two-level atom interacting with one outgoing mode of a massless scalar field in the Schwarzschild exterior. The atom follows the radial timelike trajectory established in Section 2, with conserved specific energy \(E>0\), remaining proper time \(s=\tau_H-\tau>0\), surface gravity \(\kappa=1/(2r_s)\), and outgoing retarded time \(u=t-r_*\). We retain the local outgoing geometric-optics sector appropriate to the idealized Boulware-like exterior state used in HBAR \cite{Boulware:1974dm,Scully:2017utk,Azizi:2021qcu}. The Boulware vacuum itself is not regular across the horizon, so our restriction to the exterior must not be read as a globally regular infalling field state \cite{Boulware:1974dm,Unruh:1976db}. The calculation concerns the response of a selected exterior mode and does not presume a globally normalized cavity spectrum.

\subsection{Interaction and the two transition channels}

Let \(\omega>0\) be the atomic transition frequency in proper time and \(\nu>0\) the Killing frequency of the outgoing field mode. The atomic states \(|g\rangle\) and \(|e\rangle\) differ in energy by \(\omega\). In the local outgoing approximation, we take the mode normalization \(\mathcal N_\nu\) to be positive and write the field restricted to the worldline as
\begin{equation} \widehat\Phi_\nu(\tau)=\mathcal N_\nu\left(\widehat a_\nu e^{-i\nu u(\tau)}+\widehat a_\nu^\dagger e^{i\nu u(\tau)}\right). \label{3.1} \end{equation}
The factor \(\mathcal N_\nu\) represents the chosen mode normalization and its slowly varying radial amplitude at the horizon. We hold that amplitude fixed over the local interaction region. The approximation excludes radial scattering, angular dependence, and the ingoing branch. A varying radial prefactor would generate additional corrections beyond those arising solely from the retarded-time jet. In particular, globally normalized outgoing-at-infinity modes can contain both radial-flux branches near the horizon \cite{Page:1976df,Lobos:2026ahg}.

In the interaction picture, we take a real coupling \(\lambda\) and the standard linear atom-field interaction without discarding the counter-rotating terms \cite{Scully:2003zz,Belyanin:2004cvu,Scully:2017utk},
\begin{equation} \widehat H_I(\tau)=\lambda\chi(s)\left(\widehat\sigma_+e^{i\omega\tau}+\widehat\sigma_-e^{-i\omega\tau}\right)\widehat\Phi_\nu(\tau). \label{3.2} \end{equation}
The operators \(\widehat\sigma_+=|e\rangle\langle g|\) and \(\widehat\sigma_-=|g\rangle\langle e|\) raise and lower the atomic state, respectively. We model the exterior portion of the interaction, for \(\alpha>0\) and \(0<s<\infty\), by the one-sided exponential envelope
\begin{equation} \chi(s)=e^{-\alpha s}. \label{3.3} \end{equation}
Here \(\alpha^{-1}\) gives the proper-time scale over which the integrand is appreciable before horizon crossing. The envelope ends at the horizon and is not a smooth switching function continued across it. We use it to obtain a definite local spectral problem, rather than to represent a particular physical cavity boundary. The distinction between a chosen interaction profile and a global detector response is familiar from finite-switching studies \cite{Schlicht:2003iy,Satz:2006kb,Louko:2007mu,Lobos:2026ahg}.

At first order in \(\lambda\), the counter-rotating process \(|g,n\rangle\to|e,n+1\rangle\) excites the atom while creating one photon. Its amplitude is
\begin{equation} \mathcal A_{\mathrm e}^{(n)}=-i\lambda\mathcal N_\nu\sqrt{n+1}\int_0^\infty ds\,\chi(s)e^{i\omega(\tau_H-s)}e^{i\nu u(s)}. \label{3.4} \end{equation}
For \(n\geq1\), the process \(|g,n\rangle\to|e,n-1\rangle\) excites the atom while absorbing one photon. Its amplitude is
\begin{equation} \mathcal A_{\mathrm a}^{(n)}=-i\lambda\mathcal N_\nu\sqrt n\int_0^\infty ds\,\chi(s)e^{i\omega(\tau_H-s)}e^{-i\nu u(s)}. \label{3.5} \end{equation}
Both expressions follow from the first-order Dyson term and the field ladder-operator matrix elements. Reversing the proper-time integration limits from \(\tau<\tau_H\) to \(s>0\) produces the positive integral shown. We define \(g=|\lambda|\mathcal N_\nu\) and absorb the sign of \(\lambda\) into an irrelevant constant phase. The factors \(\sqrt{n+1}\) and \(\sqrt n\) will be separated from the single-transition amplitudes, so \(P_{\mathrm e}=|\mathcal A_{\mathrm e}^{(0)}|^2\) and \(P_{\mathrm a}=|\mathcal A_{\mathrm a}^{(1)}|^2\) are the corresponding elementary probabilities.

\subsection{The affine Fourier-Mellin integral}

The leading retarded-time jet in Eq. \eqref{2.13}, retaining its first regular term, reads
\begin{equation} u_{\mathrm{lin}}(s)=u_0-\kappa^{-1}\ln(s/s_0)-2Es. \label{3.6} \end{equation}
The outgoing phase contains the power \((s/s_0)^{i\nu/\kappa}\). A translation of \(u\) therefore acts asymptotically as a dilation of \(s\), as established in Eq. \eqref{2.16}. For \(j=\mathrm e,\mathrm a\), define the dimensionless spectral parameter, shifted frequencies, and complex Laplace variables by
\begin{align} \sigma&=\frac{\nu}{\kappa},\nn \Omega_{\mathrm e}&=\omega+2E\nu,\nn \Omega_{\mathrm a}&=\omega-2E\nu,\nn z_j&=\alpha+i\Omega_j. \label{3.7} \end{align}
Substitution of Eq. \eqref{3.6} into Eq. \eqref{3.4} yields the elementary emission amplitude
\begin{equation} \mathcal A_{\mathrm e}=-ig\,e^{i\omega\tau_H+i\nu u_0}\int_0^\infty ds\,e^{-z_{\mathrm e}s}(s/s_0)^{-i\sigma}. \label{3.8} \end{equation}
The absorption phase carries the opposite imaginary power,
\begin{equation} \mathcal A_{\mathrm a}=-ig\,e^{i\omega\tau_H-i\nu u_0}\int_0^\infty ds\,e^{-z_{\mathrm a}s}(s/s_0)^{i\sigma}. \label{3.9} \end{equation}
We use the positive real logarithm to define \((s/s_0)^{\pm i\sigma}=\exp[\pm i\sigma\ln(s/s_0)]\). The same proper-time spectral measure \(ds\) occurs for both processes, while their complex powers differ by conjugation. In particular, the imaginary powers are oscillatory and have unit modulus for \(s>0\).

The relation to multiplicative harmonic analysis becomes explicit after the real change of variable \(x=\ln(s/s_0)\). For either sign, introduce
\begin{align} \mathcal I_\pm(z,\sigma)&=\int_0^\infty ds\,e^{-zs}(s/s_0)^{\pm i\sigma}\nn &=s_0\int_{-\infty}^\infty dx\,\exp\!\left[(1\pm i\sigma)x-zs_0e^x\right]. \label{3.10} \end{align}
The factor \(e^{\pm i\sigma x}\) is a Fourier phase in logarithmic proper time. The accompanying \(e^x\) originates from the Jacobian \(ds=s_0e^x dx\). In this precise sense the transition integral has a Fourier-Mellin representation adapted to the near-horizon translation-dilation correspondence. We do not identify the proper-time translation operator \(-i\partial_s\) as a self-adjoint generator on a half-line, and no such assertion is needed for the integral evaluation. Related affine-group spectral arguments appear in Refs. \cite{Arzano:2025aeo,Arzano:2026vwz}.

For \(\operatorname{Re}a>0\) and \(\operatorname{Re}z>0\), the Mellin-Laplace identity is
\begin{align} &\int_0^\infty ds\,s^{a-1}e^{-zs}=\Gamma(a)z^{-a},\nn z^{-a}&=\exp[-a\operatorname{Log}z]. \label{3.11} \end{align}
We take the principal logarithm \(\operatorname{Log}z=\ln|z|+i\arg z\), whose argument lies between \(-\pi/2\) and \(\pi/2\) for \(z=\alpha+i\Omega\) with \(\alpha>0\). Setting \(a=1\pm i\sigma\) gives the exact result for the linear-jet integrals,
\begin{equation} \mathcal I_\pm(z,\sigma)=s_0^{\mp i\sigma}\Gamma(1\pm i\sigma)z^{-1\mp i\sigma}. \label{3.12} \end{equation}
The real part of each Mellin index equals unity, and the exponential envelope ensures absolute convergence at large \(s\). The integrand is integrable at the lower endpoint because the imaginary power has unit magnitude. No unspecified boundary term enters the transform.

\subsection{Emission, absorption, and the spectral ratio}

Using Eq. \eqref{3.12} in Eq. \eqref{3.8}, we obtain
\begin{equation} \mathcal A_{\mathrm e}=-ig\,e^{i\omega\tau_H+i\nu u_0}s_0^{i\sigma}\Gamma(1-i\sigma)z_{\mathrm e}^{-1+i\sigma}. \label{3.13} \end{equation}
The corresponding absorption amplitude is
\begin{equation} \mathcal A_{\mathrm a}=-ig\,e^{i\omega\tau_H-i\nu u_0}s_0^{-i\sigma}\Gamma(1+i\sigma)z_{\mathrm a}^{-1-i\sigma}. \label{3.14} \end{equation}
The proper-time and retarded-time reference phases have unit magnitude and cancel separately in the probabilities. Since \(\nu>0\) and \(\kappa>0\), we have \(\sigma>0\). Complex conjugation of the gamma function, followed by Euler's reflection identity, yields
\begin{equation} |\Gamma(1+i\sigma)|^2=|\Gamma(1-i\sigma)|^2=\frac{\pi\sigma}{\sinh(\pi\sigma)}. \label{3.15} \end{equation}
To evaluate the remaining complex powers, write, for \(j=\mathrm e,\mathrm a\),
\begin{align} \rho_j&=(\alpha^2+\Omega_j^2)^{1/2},\nn \theta_j&=\arg z_j=\arctan(\Omega_j/\alpha). \label{3.16} \end{align}
The emission power contains the factor \(\rho_{\mathrm e}^{-2}e^{-2\sigma\theta_{\mathrm e}}\). The absorption power instead contains \(\rho_{\mathrm a}^{-2}e^{2\sigma\theta_{\mathrm a}}\). The elementary probabilities therefore read
\begin{equation} P_{\mathrm e}=\frac{g^2\pi\sigma}{\sinh(\pi\sigma)}\frac{e^{-2\sigma\theta_{\mathrm e}}}{\alpha^2+\Omega_{\mathrm e}^2}. \label{3.17} \end{equation}
\begin{equation} P_{\mathrm a}=\frac{g^2\pi\sigma}{\sinh(\pi\sigma)}\frac{e^{2\sigma\theta_{\mathrm a}}}{\alpha^2+\Omega_{\mathrm a}^2}. \label{3.18} \end{equation}
Taking the quotient eliminates the common coupling and gamma-function modulus, leaving the spectral detailed-balance ratio
\begin{equation} q(\nu;\alpha,E)=\frac{P_{\mathrm e}}{P_{\mathrm a}}=\frac{\alpha^2+\Omega_{\mathrm a}^2}{\alpha^2+\Omega_{\mathrm e}^2}e^{-2\sigma(\theta_{\mathrm e}+\theta_{\mathrm a})}. \label{3.19} \end{equation}
The ratio is exact for the specified exponential envelope and the retarded-time phase truncated after its linear regular term. In the branch \(\omega>2E\nu\), both shifted frequencies are positive and the expression obeys \(0<q<1\). At this stage, no claim of exact thermal detailed balance is required. The dependence on \(\alpha\) and \(E\) remains explicit and permits subsequent asymptotic analysis.
The fixed-amplitude assumption also has a definite spectral consequence. Suppose a real radial prefactor varies as \(\mathcal R_\nu(s)=\mathcal R_{\nu,H}[1+a_1s+\mathcal{O}(s^2)]\), with \(a_1\in\mathbb R\) and dimensions of inverse proper time. Multiplying the transition integrands by the linear term \(a_1s\) and evaluating the same Mellin-Laplace integrals gives the relative amplitude corrections \(\delta_{\mathrm e,\mathrm{amp}}=a_1(1-i\sigma)/z_{\mathrm e}\) and \(\delta_{\mathrm a,\mathrm{amp}}=a_1(1+i\sigma)/z_{\mathrm a}\). These terms are generally of order \(\omega^{-1}\), whereas the quadratic phase correction is of order \(\omega^{-2}\) at fixed \(\sigma\) and \(\nu\). The isolated spherical factor \(r_s/r(s)\) gives \(a_1=-E/r_s\), since \(r(s)=r_s+Es+\mathcal{O}(s^2)\). At first order, its contribution to the logarithmic quotient is \(2\operatorname{Re}(\delta_{\mathrm e,\mathrm{amp}}-\delta_{\mathrm a,\mathrm{amp}})\). In the formal Abel limit \(\alpha\to0^+\), with \(\omega>2E\nu\), this becomes \(-2a_1\sigma(\Omega_{\mathrm e}^{-1}+\Omega_{\mathrm a}^{-1})\). For \(a_1=-E/r_s\), it gives \(+8E\nu/\omega+\mathcal{O}(\omega^{-3})\) at high gap. The phase-only quotient has the opposite term \(-8E\nu/\omega+\mathcal{O}(\omega^{-3})\), so the two contributions cancel at this order for this factor. The cancellation is not a result for a complete scattering mode. Radial potential scattering and the ingoing branch can add further frequency-dependent structure, and the present spectral corrections therefore remain phase contributions within the frozen-amplitude approximation.

The gamma-function expression follows from extending the linear retarded-time jet over the full half-line in a model integral. Its relation to the exact exterior trajectory requires localization. For fixed \(E>0\), the relevant part of the interaction must satisfy \(s/r_s\ll1\) and \(f(r(s))\ll E^2\). At the nominal exponential-envelope scale \(s=1/\alpha\), the leading geodesic expansion gives \((r(s)-r_s)/r_s\simeq E/(\alpha r_s)\) and \(f(r(s))/E^2\simeq1/(E\alpha r_s)\). A sufficient finite-parameter hierarchy for arbitrary fixed \(E>0\) is therefore \(\alpha r_s\gg\max\{E,E^{-1}\}\). For field frequencies of order \(\kappa\), the first omitted phase term in Eq. \eqref{2.15} has magnitude of order \(\nu s^2/r_s\), so its smallness over the nominal envelope scale also requires \(\nu/(r_s\alpha^2)\ll1\). The error in an oscillatory transition integral depends on endpoint cancellations and on radial variation of the field mode. The gamma-function formulas are exact identities for the specified local model, not uniform error bounds for the global Schwarzschild response.

The affine spectral representation separates the origin of the near-horizon imaginary power from the regular Doppler-like frequency shifts and the interaction envelope. The same integral identity determines both atomic transition channels, and the unequal complex-power phases generate the nontrivial quotient in Eq. \eqref{3.19}. Neither the use of the gamma function nor the translation-dilation correspondence is new by itself \cite{Scully:2017utk,Arzano:2025aeo,Arzano:2026vwz}. The present formulation provides a definite starting point for examining the high-gap HBAR limit and the corrections associated with successive worldline jets.

\section{Thermal Detailed Balance and the HBAR Limit}
\label{sec4}
The affine spectral amplitudes obtained in Section 3 allow us to identify the circumstances under which a freely falling atom produces a thermal emission-to-absorption ratio. We retain the Schwarzschild geometry, the outgoing Boulware-like mode sector, and the first-order two-level atom interaction specified previously. In geometric units, the atomic gap is \(\omega>0\), the outgoing Killing frequency is \(\nu>0\), the surface gravity is \(\kappa=1/(2r_s)\), and \(\sigma=\nu/\kappa\). The geodesic energy \(E>0\) remains arbitrary but fixed. We distinguish exact identities for the linear-jet spectral model from asymptotic statements about the underlying Schwarzschild problem.

\subsection{The Abel limit and the Planck factor}

The single-photon probabilities in Eqs. \eqref{3.17} and \eqref{3.18} have a common gamma-function modulus but opposite phases of the complex Laplace variables. Their ratio is
\begin{equation} q(\nu;\alpha,E)=\frac{P_{\mathrm e}}{P_{\mathrm a}}=\frac{\alpha^2+\Omega_{\mathrm a}^2}{\alpha^2+\Omega_{\mathrm e}^2}\exp[-2\sigma(\theta_{\mathrm e}+\theta_{\mathrm a})]. \label{4.1} \end{equation}
Here \(\theta_j=\arctan(\Omega_j/\alpha)\) for \(\alpha>0\). The gamma-function factor itself is
\begin{equation} |\Gamma(1+i\sigma)|^2=\frac{\pi\sigma}{\sinh(\pi\sigma)}. \label{4.2} \end{equation}
Its combination with the phase acquired by the emission amplitude yields the identity
\begin{equation} \frac{\pi\sigma e^{-\pi\sigma}}{\sinh(\pi\sigma)}=\frac{2\pi\sigma}{e^{2\pi\sigma}-1}. \label{4.3} \end{equation}
The Planck denominator therefore follows from the gamma-function integral and its complex phase, rather than from an imposed thermal population \cite{Scully:2017utk,Camblong:2020pme}.

First consider the Abel boundary value \(\alpha\to0^+\) of the exact linear-jet expressions. We restrict attention to \(\omega>2E\nu\), so both shifted frequencies are positive and the chosen complex-power branch is unambiguous. The limiting arguments satisfy
\begin{equation} \lim_{\alpha\to0^+}\theta_{\mathrm e}=\lim_{\alpha\to0^+}\theta_{\mathrm a}=\frac{\pi}{2}. \label{4.4} \end{equation}
Consequently, the elementary probabilities become
\begin{equation} P_{\mathrm e}^{(0)}=\frac{2\pi g^2\sigma}{(\omega+2E\nu)^2}\frac{1}{e^{2\pi\sigma}-1}. \label{4.5} \end{equation}
\begin{equation} P_{\mathrm a}^{(0)}=\frac{2\pi g^2\sigma}{(\omega-2E\nu)^2}\frac{e^{2\pi\sigma}}{e^{2\pi\sigma}-1}. \label{4.6} \end{equation}
The common factor \(g^2\) contains the fixed local mode amplitude and the atom-field coupling defined in Section 3. Dividing gives
\begin{equation} q_0(\nu;E)=\left(\frac{\omega-2E\nu}{\omega+2E\nu}\right)^2e^{-2\pi\nu/\kappa}. \label{4.7} \end{equation}
The exponential is already of thermal form, whereas the rational factor shows that finite-gap detailed balance is not exactly thermal even in this Abel limit. In particular, the probabilities in Eqs. \eqref{4.5} and \eqref{4.6} are not equal to equilibrium photon occupation numbers.

For an atom released from rest at infinity, \(E=1\). In the usual \(r_s=1\) dimensionless Schwarzschild convention, \(\kappa=1/2\), and Eq. \eqref{4.5} reads
\begin{equation} P_{\mathrm e}^{(0)}=\frac{4\pi g^2\nu}{\omega^2(1+2\nu/\omega)^2}\frac{1}{e^{4\pi\nu}-1}. \label{4.8} \end{equation}
The expression agrees algebraically with the leading high-gap excitation probability derived by Scully and collaborators in their Appendix B, including the \((1+2\nu/\omega)^{-2}\) prefactor \cite{Scully:2017utk}. That agreement refers to the corresponding outgoing-mode normalization and asymptotic regime. It does not establish equality of the complete global field responses at finite atomic gap.

\subsection{High-gap detailed balance with a localized interaction}

The Abel limit is an identity of the extended half-line spectral integral. It does not justify replacing the exact exterior trajectory by a near-horizon jet at arbitrarily large remaining proper time. A physically localized asymptotic regime retains \(\alpha>0\) and requires the interaction region to remain near the horizon. We retain the finite-parameter localization hierarchy \(\alpha r_s\gg\max\{E,E^{-1}\}\) established in Section 3. We also require a high atomic gap with \(\omega\gg\alpha\) and \(\omega\gg2E\nu\). We take \(\sigma=\nu/\kappa\) in a fixed bounded positive range. These conditions allow the switching scale to remain local while the atomic gap becomes large.

Since \(\omega>2E\nu\) and \(\alpha>0\), these ratios satisfy \(0<\eta<1\) and \(\delta>0\). Introduce the dimensionless ratios and their maximum by
\begin{align} \eta&=\frac{2E\nu}{\omega},\nn \delta&=\frac{\alpha}{\omega}, \label{4.9} \end{align}
with $\varepsilon=\max\{\eta,\delta\}$. Both \(\eta\) and \(\delta\) tend to zero in the stated high-gap regime. Since \(\Omega_{\mathrm e}\) and \(\Omega_{\mathrm a}\) remain positive, the identity \(\arctan(x/\delta)=\pi/2-\arctan(\delta/x)\) applies for their positive dimensionless values. Equation \eqref{4.1} therefore becomes the exact logarithmic relation
\begin{align} \ln q&=-2\pi\sigma+\ln\!\left[\frac{(1-\eta)^2+\delta^2}{(1+\eta)^2+\delta^2}\right]
\nn &+2\sigma\left[\arctan\!\left(\frac{\delta}{1+\eta}\right)+\arctan\!\left(\frac{\delta}{1-\eta}\right)\right]. \label{4.10} \end{align}
Taylor expansion about \((\eta,\delta)=(0,0)\) gives the explicit leading correction
\begin{equation} \ln q=-2\pi\sigma-4\eta+4\sigma\delta+\mathcal{O}_{\sigma}(\varepsilon^3). \label{4.11} \end{equation}
The estimate holds uniformly for \(\sigma\) in any fixed compact interval and \(\varepsilon\) sufficiently small. The remainder begins at cubic total degree because the logarithm of the rational factor is odd in \(\eta\), while the arctangent sum contributes a linear term in \(\delta\) followed by terms of cubic or higher total degree. The constants implicit in the expansion need not remain bounded when \(\sigma\) increases without bound.

Exponentiating Eq. \eqref{4.11} yields
\begin{equation} q=e^{-2\pi\sigma}\left[1-4\eta+4\sigma\delta+\mathcal{O}_{\sigma}(\varepsilon^2)\right]. \label{4.12} \end{equation}
The emission and absorption probabilities separately admit high-gap expansions obtained directly from Eqs. \eqref{3.17} and \eqref{3.18},
\begin{equation} P_{\mathrm e}=\frac{2\pi g^2\sigma}{\omega^2(e^{2\pi\sigma}-1)}\left[1-2\eta+2\sigma\delta+\mathcal{O}_{\sigma}(\varepsilon^2)\right]. \label{4.13} \end{equation}
\begin{equation} P_{\mathrm a}=\frac{2\pi g^2\sigma e^{2\pi\sigma}}{\omega^2(e^{2\pi\sigma}-1)}\left[1+2\eta-2\sigma\delta+\mathcal{O}_{\sigma}(\varepsilon^2)\right]. \label{4.14} \end{equation}
The three expansions are mutually consistent upon division, and the correction terms distinguish the regular worldline phase from the interaction envelope. The common high-gap limit follows without sending \(\alpha\) to zero. At fixed \(E\) and bounded \(\sigma\), with \(\eta,\delta\to0\), we obtain
\begin{equation} \lim_{\eta,\delta\to0}q(\nu;\alpha,E)=e^{-2\pi\nu/\kappa}. \label{4.15} \end{equation}
The limit denotes the stated double scale separation, not an unrestricted limit of a finite-energy atom or an exact finite-duration Kubo-Martin-Schwinger property of the field state. The leading ratio is independent of \(E\), although the finite-gap prefactors and the first corrections retain its influence.

\subsection{Hawking temperature and physical interpretation}

A bosonic mode of energy \(\nu\) at inverse temperature \(\beta\) has thermal detailed-balance factor \(e^{-\beta\nu}\) in geometric units. Comparison with Eq. \eqref{4.15} identifies the inverse temperature and the associated temperature as
\begin{align} \beta_H&=\frac{2\pi}{\kappa},\nn T_H=\frac{\kappa}{2\pi}&=\frac{1}{8\pi M}. \label{4.16} \end{align}
The parameter \(\nu\) is the Killing frequency of the selected outgoing mode, so this temperature refers to the normalization of time for a static observer at infinity. Returning to physical units, let \(\kappa_{\mathrm{freq}}=c^3/(4GM)\) denote the surface-gravity frequency scale. Then
\begin{equation} T_H=\frac{\hbar\kappa_{\mathrm{freq}}}{2\pi k_B}=\frac{\hbar c^3}{8\pi GMk_B}. \label{4.17} \end{equation}
The temperature equals the Schwarzschild Hawking temperature \cite{Hawking:1975vcx,Unruh:1976db}, but equality of the leading transition ratio does not identify the emitted HBAR photons with spontaneous Hawking quanta \cite{Scully:2017utk,Azizi:2021qcu}. The former arise from transitions of injected atoms in a specified exterior field state, whereas the latter do not require the atomic medium. Likewise, the Planck denominator in Eq. \eqref{4.5} does not alone establish an equilibrium density matrix for the photon field. That additional statement requires dynamical assumptions about repeated atomic injections and radiative occupations.

The atom itself travels on a geodesic. If \(U^\mu=dx^\mu/d\tau\) denotes its four-velocity, its proper four-acceleration satisfies
\begin{equation} a^\mu=U^\nu\nabla_\nu U^\mu=0. \label{4.18} \end{equation}
The exponential in Eq. \eqref{4.15} thus does not arise from assigning uniform proper acceleration to the atom. Its mathematical origin is the coefficient \(-1/\kappa\) of the logarithm in the outgoing retarded-time jet of Eq. \eqref{2.13}. The Fourier phase of the external mode becomes an imaginary power of the remaining proper time, and the opposite complex-power branches in emission and absorption generate the relative factor \(e^{-2\pi\sigma}\). Curvature, field-state choice, and the atom-mode relative evolution are essential to that interpretation.

The transition quotient and its high-gap limit apply to the selected local outgoing mode with frozen radial amplitude and the specified interaction envelope. They do not determine a complete Schwarzschild radiation flux, for which global mode normalization, radial scattering, and ingoing components would also have to be included. The finite-parameter expression in Eq. \eqref{4.1} remains the starting point for departures from detailed balance within the present model.

\section{Higher-Order Worldline Jets and Spectral Departures from Thermality}
\label{sec5}

The affine spectral amplitudes of Section 3 reproduce the leading HBAR detailed-balance factor in the high-gap regime, while retaining finite-gap and finite-envelope corrections. We now examine how the first nonlinear term of the infalling trajectory modifies those amplitudes. We also characterize the frequency dependence of the logarithmic emission-to-absorption ratio. Throughout, we use the Schwarzschild geometry and the local outgoing scalar-mode approximation introduced in Sections 2 and 3. The quantities \(r_s\), \(\kappa=1/(2r_s)\), \(E>0\), \(s=\tau_H-\tau\), \(\omega>0\), \(\nu>0\), and \(\alpha>0\) retain their previous meanings. We distinguish identities of the idealized spectral integrals from asymptotic claims about a localized segment of the exact geodesic.

\subsection{Successive worldline coefficients and Mellin moments}

The radial geodesic equation in Eq. \eqref{2.3} and the identity \(r''=-r_s/(2r^2)\) determine further derivatives at horizon crossing. In particular, differentiating twice more gives \(r''''(0)=-(3E^2+1/2)/r_s^3\). With \(y=s/r_s\), the radial series through fourth degree is
\begin{equation} \frac{r(s)}{r_s}=1+Ey-\frac{y^2}{4}+\frac{E}{6}y^3-\left(\frac{E^2}{8}+\frac{1}{48}\right)y^4+\mathcal{O}_E(y^5). \label{5.1} \end{equation}
Substitution into the outgoing-null relation of Eq. \eqref{2.9} gives \(du/ds=-2r_s/s-2E+s/(3r_s)-Es^2/(4r_s^2)+\mathcal{O}_E(s^3/r_s^3)\). Integrating, with the same additive reference as in Section 2, yields
\begin{equation} u(s)=u_0-\kappa^{-1}\ln(s/s_0)-2Es+\frac{s^2}{6r_s}-\frac{Es^3}{12r_s^2}+\mathcal{O}_E\!\left(\frac{s^4}{r_s^3}\right). \label{5.2} \end{equation}
The first nonlinear correction to the linear retarded-time approximation is therefore independent of \(E\). The following cubic coefficient depends on \(E\), and its explicit value will delimit the accuracy of a quadratic truncation. The remainder is local for each fixed \(E>0\), without uniformity as \(E\to0^+\).

Define the quadratic retarded-time jet by
\begin{equation} u_{[2]}(s)=u_0-\kappa^{-1}\ln(s/s_0)-2Es+\frac{s^2}{6r_s}. \label{5.3} \end{equation}
For the photon-emission branch, the extra factor relative to the linear model is \(\exp[i\nu s^2/(6r_s)]\), whereas the absorption branch carries its complex conjugate. At sufficiently small positive \(s\), these factors obey
\begin{align} e^{\pm i\nu s^2/(6r_s)}&=1\pm\frac{i\nu s^2}{6r_s}+R_{\pm}(s),\nn |R_{\pm}(s)|&\leq\frac{\nu^2s^4}{72r_s^2}. \label{5.4} \end{align}
The bound follows from the integral remainder for an exponential with real argument. In addition, the phase difference between the exact worldline and \(u_{[2]}\) is bounded locally by \(\nu C_Es^3/r_s^2\), with \(C_E\) finite on a sufficiently small exterior interval.

The new factor \(s^2\) does not require a different spectral transformation. Let \(n\) be a nonnegative integer and let \(\operatorname{Re}z>0\). The higher Mellin-Laplace moments are
\begin{eqnarray}
    \mathcal M_{\pm,n}(z,\sigma)&=&\int_0^\infty ds\,s^n e^{-zs}(s/s_0)^{\pm i\sigma}=\\\nonumber
&=&s_0^{\mp i\sigma}\Gamma(n+1\pm i\sigma)z^{-n-1\mp i\sigma}. \label{5.5} 
\end{eqnarray}
The complex powers use the principal logarithm specified in Section 3. Dividing by the zeroth moment gives the finite-product rule
\begin{align} \frac{\mathcal M_{\pm,n}(z,\sigma)}{\mathcal M_{\pm,0}(z,\sigma)}&=\frac{\prod_{m=1}^{n}(m\pm i\sigma)}{z^n},\nn \mathcal M_{\pm,n}&=(-\partial_z)^n\mathcal M_{\pm,0}. \label{5.6} \end{align}
For \(n=0\), the product is defined as unity. The derivative identity also shows that successive powers of the proper-time coordinate can be introduced by derivatives of the same spectral integral. The result follows directly from the gamma-function recurrence and is valid for either transition branch.

Write \(\mathcal A_j^{\mathrm{lin}}\) for the linear-jet amplitudes in Eqs. \eqref{3.13} and \eqref{3.14}, with \(j=\mathrm e,\mathrm a\), and write \(P_j^{\mathrm{lin}}=|\mathcal A_j^{\mathrm{lin}}|^2\) for their probabilities. We retain \(\sigma=\nu/\kappa\), \(z_{\mathrm e}=\alpha+i(\omega+2E\nu)\), and \(z_{\mathrm a}=\alpha+i(\omega-2E\nu)\). The first quadratic spectral approximation has the form
\begin{align} \widetilde{\mathcal A}_{\mathrm e}^{[2]}&=\mathcal A_{\mathrm e}^{\mathrm{lin}}(1+\delta_{\mathrm e}^{(2)}),\nn \widetilde{\mathcal A}_{\mathrm a}^{[2]}&=\mathcal A_{\mathrm a}^{\mathrm{lin}}(1+\delta_{\mathrm a}^{(2)}). \label{5.7} \end{align}
The moment of degree two fixes the emission correction without any undetermined numerical coefficient,
\begin{equation} \delta_{\mathrm e}^{(2)}=\frac{i\nu}{6r_s}\frac{(1-i\sigma)(2-i\sigma)}{z_{\mathrm e}^2}. \label{5.8} \end{equation}
The absorption correction follows from the opposite field phase,
\begin{equation} \delta_{\mathrm a}^{(2)}=-\frac{i\nu}{6r_s}\frac{(1+i\sigma)(2+i\sigma)}{z_{\mathrm a}^2}. \label{5.9} \end{equation}
All constant reference phases remain in \(\mathcal A_j^{\mathrm{lin}}\) and disappear in the probabilities. Using the shifted frequencies from Eq. \eqref{3.7}, a sufficient condition for retaining only the displayed quadratic contribution in each amplitude is
\begin{equation} |\delta_j^{(2)}|=\frac{\nu\sqrt{(1+\sigma^2)(4+\sigma^2)}}{6r_s(\alpha^2+\Omega_j^2)}\ll1. \label{5.10} \end{equation}
Smallness of this term alone does not bound the influence of the omitted cubic worldline coefficient or the remote portion of a physical interaction. Both require a separate locality argument.

We make that argument explicit by introducing a smooth auxiliary envelope \(h(s)\) with \(0\leq h\leq1\), \(h=1\) for \(0\leq s\leq L\), and \(h=0\) for \(s\geq2L\). Choose \(2L/r_s\ll1\) with \(f(r(2L))\ll E^2\). Suppose \(|u(s)-u_{[2]}(s)|\leq C_Es^3/r_s^2\) on its support. Let \(\mathcal A_j^{h}\) denote the corresponding exact-worldline amplitude with envelope \(h(s)e^{-\alpha s}\) and the same constant mode prefactor \(g\). Direct integration of the absolute phase-remainder bounds yields
\begin{equation} \left|\mathcal A_j^{h}-\widetilde{\mathcal A}_j^{[2]}\right|\leq g\left[\frac{6C_E\nu}{r_s^2\alpha^4}+\frac{\nu^2}{3r_s^2\alpha^5}+\frac{e^{-\alpha L}}{\alpha}\right]. \label{5.11} \end{equation}
The three terms arise, respectively, from the first omitted worldline power, the second-order remainder of the quadratic phase, and the difference between the smooth local envelope and the extended half-line integral. The bound is conservative because it ignores oscillatory cancellations. It becomes useful for a probability ratio only when its right-hand side is small compared with the magnitudes of both unperturbed amplitudes.

An additional high-gap statement is available for fixed \(E\), \(r_s\), \(\alpha\), \(\nu\), and smooth \(h\), with \(\omega\to\infty\) and \(\omega>2E\nu\). The generalized Fourier endpoint expansion of the smooth regular phase multiplying \(s^{\mp i\sigma}\) gives
\begin{equation} \frac{\mathcal A_j^{h}}{\mathcal A_j^{\mathrm{lin}}}=1+\delta_j^{(2)}+\mathcal{O}_{E,\sigma,\alpha,L}\!\left(\frac{\nu}{r_s^2\omega^3}+\frac{\nu^2}{r_s^2\omega^4}\right). \label{5.12} \end{equation}
The statement holds with \(\sigma\) in a fixed bounded interval and with a smooth local envelope equal to unity near the endpoint. Away from \(s=0\), repeated integration by parts gives terms that fall faster than any prescribed inverse power of \(\omega\). The estimate is not asserted uniformly when \(\alpha\), \(\nu\), or \(E\) scale with the atomic gap. In the overlapping regime in which the exponential envelope itself localizes the interaction, we retain the hierarchy stated in Section 3 and additionally require \(\omega\gg\alpha\).

\subsection{Probability corrections and the nonthermal ratio}

Assume the correction magnitudes in Eq. \eqref{5.10} are small and the remainder conditions above are satisfied. For either transition channel \(j=\mathrm e,\mathrm a\), squaring the perturbative amplitudes gives
\begin{equation} \widetilde P_j^{[2]}=P_j^{\mathrm{lin}}\left[1+2\operatorname{Re}\delta_j^{(2)}+|\delta_j^{(2)}|^2\right]. \label{5.13} \end{equation}
For the emission-to-absorption quotient, the first nonvanishing quadratic-jet correction is
\begin{align} \widetilde q_{[2]}&=q_{\mathrm{lin}}\left[1+\Delta_2(\nu)+\mathcal{O}\!\left(|\delta_{\mathrm e}^{(2)}|^2+|\delta_{\mathrm a}^{(2)}|^2\right)\right],\nn \Delta_2&=2\operatorname{Re}(\delta_{\mathrm e}^{(2)}-\delta_{\mathrm a}^{(2)}). \label{5.14} \end{align}
Here \(q_{\mathrm{lin}}\) is the exact linear-jet ratio of Eq. \eqref{3.19}. The tilde signals that \(\widetilde q_{[2]}\) is defined using the first quadratic spectral amplitudes, not the exact probabilities from the complete Schwarzschild trajectory.

For an explicitly real formula, define \(R_j=\alpha^2+\Omega_j^2\) and \(B=2-\sigma^2\). The emission correction has real part
\begin{equation} \operatorname{Re}\delta_{\mathrm e}^{(2)}=\frac{\nu}{6r_s}\frac{2B\alpha\Omega_{\mathrm e}+3\sigma(\alpha^2-\Omega_{\mathrm e}^2)}{R_{\mathrm e}^2}. \label{5.15} \end{equation}
The corresponding absorption term is
\begin{equation} \operatorname{Re}\delta_{\mathrm a}^{(2)}=\frac{\nu}{6r_s}\frac{-2B\alpha\Omega_{\mathrm a}+3\sigma(\alpha^2-\Omega_{\mathrm a}^2)}{R_{\mathrm a}^2}. \label{5.16} \end{equation}
Introduce two rational functions, whose sign difference preserves the distinction between emission and absorption,
\begin{equation} F_{\pm}(x)=\frac{\pm2B\alpha x+3\sigma(\alpha^2-x^2)}{(\alpha^2+x^2)^2}. \label{5.17} \end{equation}
The resulting frequency-dependent departure from the linear-jet probability ratio is
\begin{equation} \Delta_2(\nu)=\frac{\nu}{3r_s}\left[F_+(\Omega_{\mathrm e})-F_-(\Omega_{\mathrm a})\right]. \label{5.18} \end{equation}
A useful analytic check is the Abel boundary value for \(\omega>2E\nu\). Taking \(\alpha\to0^+\) in the quadratic coefficient gives
\begin{equation} \Delta_2(\nu)\big|_{\alpha\to0^+}=\frac{8E\omega\nu^2\sigma}{r_s(\omega^2-4E^2\nu^2)^2}. \label{5.19} \end{equation}
For positive \(E\), \(\nu\), and \(\omega>2E\nu\), the right-hand side is positive. It is a formal limiting value of the local spectral coefficient rather than a justification for using a quadratic near-horizon jet at arbitrarily large proper time. In the large-gap regime it is suppressed by three inverse powers of \(\omega\) at fixed \(\nu\), consistently with the expected separation between the thermal leading term and higher worldline effects.

\subsection{Spectral curvature and the range of the thermal interpretation}

The finite-envelope linear-jet ratio already has a nontrivial frequency dependence. Define its negative logarithm, with the angular variables of Eq. \eqref{3.16}, by
\begin{equation} \Lambda_{\mathrm{lin}}(\nu)=-\ln q_{\mathrm{lin}}(\nu)=\ln\!\left(\frac{R_{\mathrm e}}{R_{\mathrm a}}\right)+\frac{2\nu}{\kappa}(\theta_{\mathrm e}+\theta_{\mathrm a}). \label{5.20} \end{equation}
For fixed \(\alpha>0\), \(\omega>0\), \(E>0\), and \(\kappa>0\), the expression is analytic near \(\nu=0\). Interchanging \(\nu\) and \(-\nu\) interchanges the shifted frequencies while reversing \(\sigma\). Hence \(\Lambda_{\mathrm{lin}}\) is odd, and its small-frequency series reads
\begin{align} \Lambda_{\mathrm{lin}}(-\nu)&=-\Lambda_{\mathrm{lin}}(\nu),\nn \Lambda_{\mathrm{lin}}(\nu)&=b_1^{\mathrm{lin}}\nu+b_3^{\mathrm{lin}}\nu^3+\mathcal{O}(\nu^5). \label{5.21} \end{align}
The first coefficient is
\begin{equation} b_1^{\mathrm{lin}}=\frac{4}{\kappa}\arctan\!\left(\frac{\omega}{\alpha}\right)+\frac{8E\omega}{\alpha^2+\omega^2}. \label{5.22} \end{equation}
Writing \(R_0=\alpha^2+\omega^2\), direct third-order differentiation yields
\begin{equation} b_3^{\mathrm{lin}}=-\frac{16E^2\alpha\omega}{\kappa R_0^2}+\frac{32E^3\omega(\omega^2-3\alpha^2)}{3R_0^3}. \label{5.23} \end{equation}
Both formulas are exact derivatives of the linear-jet model, with no high-gap expansion used to obtain them. For generic positive parameters, \(b_3^{\mathrm{lin}}\neq0\). The resulting ratio therefore cannot have the form \(e^{-\beta\nu}\) with a single constant \(\beta\) throughout a neighborhood of zero frequency. A vanishing cubic term at special parameter values would not by itself establish exact thermality, since higher odd coefficients may remain nonzero.

The quadratic trajectory coefficient further changes the logarithmic ratio. To first order in the quadratic spectral moments, we find
\begin{equation} -\ln\widetilde q_{[2]}(\nu)=\Lambda_{\mathrm{lin}}(\nu)-\Delta_2(\nu)+\mathcal{O}\!\left(|\delta_{\mathrm e}^{(2)}|^2+|\delta_{\mathrm a}^{(2)}|^2\right). \label{5.24} \end{equation}
The linear and cubic coefficients obtained from the first-order term \(-\Delta_2\) may be written as \(\delta b_1^{(2)}\) and \(\delta b_3^{(2)}\). Differentiation of Eq. \eqref{5.18} gives
\begin{equation} \delta b_1^{(2)}=-\frac{8\alpha\omega}{3r_sR_0^2}. \label{5.25} \end{equation}
At third order the result is
\begin{align} \delta b_3^{(2)}&=\frac{4\omega}{3\kappa^2r_sR_0^4}\Biggl[\alpha R_0^2+6E\kappa R_0(3\alpha^2-\omega^2) \nn &+48E^2\alpha\kappa^2(\alpha^2-\omega^2)\Biggr]. \label{5.26} \end{align}
The superscript \((2)\) on these coefficients identifies the contribution linear in the quadratic worldline phase. Neither expression is the full frequency-expansion coefficient of the exact Schwarzschild response. Terms quadratic in \(\delta_j^{(2)}\), and contributions from higher worldline powers, also enter the logarithm at sufficiently high frequency order.

The next geometric coefficient in Eq. \eqref{5.2} makes that limitation concrete. Its degree-three Mellin moment gives the additional first-order relative amplitudes
\begin{align} \delta_{\mathrm e}^{(3)}&=-\frac{iE\nu}{12r_s^2}\frac{\prod_{m=1}^{3}(m-i\sigma)}{z_{\mathrm e}^3},\nn \delta_{\mathrm a}^{(3)}&=\frac{iE\nu}{12r_s^2}\frac{\prod_{m=1}^{3}(m+i\sigma)}{z_{\mathrm a}^3}. \label{5.27} \end{align}
For fixed \(\sigma\), the absolute sizes of these terms decrease with an additional inverse power of the atomic gap relative to the quadratic amplitudes. Nevertheless, the third derivative of a spectral ratio with respect to \(\nu\) can receive contributions from their explicit \(\sigma\) dependence. We therefore use Eqs. \eqref{5.25} and \eqref{5.26} to identify a specific quadratic-jet contribution, not to claim that it exhausts the spectral curvature of the full worldline. This distinction is particularly important when comparing asymptotic expansions taken in the atomic gap with Taylor expansions taken in the outgoing photon frequency.

For the linear-jet model alone, the quantity \(\beta_{\mathrm{eff}}(\nu)=\Lambda_{\mathrm{lin}}(\nu)/\nu\) is a convenient way of expressing the frequency dependence for positive \(\nu\). Near zero frequency it obeys
\begin{align} \beta_{\mathrm{eff}}(\nu)&=b_1^{\mathrm{lin}}+b_3^{\mathrm{lin}}\nu^2+\mathcal{O}(\nu^4),\nn \lim_{\omega\to\infty}b_1^{\mathrm{lin}}&=\frac{2\pi}{\kappa},\nn \lim_{\omega\to\infty}b_3^{\mathrm{lin}}&=0. \label{5.28} \end{align}
The limits hold at fixed \(E\), \(\alpha\), and \(\kappa\). The symbol \(\beta_{\mathrm{eff}}\) represents the slope inferred from an individual transition ratio. When it depends on \(\nu\), it does not establish a field in thermal equilibrium or a genuine mode-independent temperature.

The formulas above give the linear- and quadratic-jet spectral contributions for the one-sided envelope and frozen radial amplitude. The nonzero cubic coefficient excludes a single frequency-independent inverse temperature within the finite-parameter linear-jet model. The coefficients \(\delta b_1^{(2)}\) and \(\delta b_3^{(2)}\) isolate the first-order effect of the quadratic phase, while higher worldline moments and radial-mode variations remain outside that partial correction.

\section{Photon Statistics and the Conditional HBAR Entropy-Area Relation}
\label{sec6}

The transition probabilities obtained in Sections 3-5 refer to individual atom-mode interactions. A statement about the entropy of the radiation requires an additional dynamical description of repeated interactions and of the resulting photon-number distribution. We adopt the dilute, randomly injected atomic beam considered in HBAR quantum optics \cite{PhysRevLett.16.853,Scully:2017utk,Azizi:2021qcu,Azizi:2021yto}. The field is restricted to the independent outgoing modes and the weak-coupling, random-phase regime in which the coarse-grained density matrix is diagonal in the photon-number basis. The following analysis first establishes the stationary distribution, then derives an exact entropy balance for the birth-death process, and finally identifies the assumptions needed to associate its radiative energy flux with an area quantity. We continue to use \(G=c=\hbar=k_B=1\) unless physical constants are explicitly restored.

\subsection{Atomic injection and the photon-number distribution}

Let \(\mathcal J>0\) be the mean number of atoms injected per unit Schwarzschild Killing time in the stationary exterior description. We reserve \(\kappa\) for the surface gravity. Each atom begins in the same ground state, and random arrival times eliminate the relevant phase coherences after coarse graining. For a selected outgoing mode of Killing frequency \(\nu>0\), the elementary probabilities \(P_{\mathrm e,\nu}\) and \(P_{\mathrm a,\nu}\) are those defined in Section 3, or their explicitly stated higher-jet approximations from Section 5. At leading order in the weak atom-field coupling, the corresponding elementary rates are
\begin{align} \Gamma_{\mathrm e,\nu}&=\mathcal J P_{\mathrm e,\nu},\nn \Gamma_{\mathrm a,\nu}&=\mathcal J P_{\mathrm a,\nu},\nn q_\nu&=\frac{\Gamma_{\mathrm e,\nu}}{\Gamma_{\mathrm a,\nu}}=\frac{P_{\mathrm e,\nu}}{P_{\mathrm a,\nu}}. \label{6.1} \end{align}
The first-order description requires the probabilities of the relevant number-enhanced transitions to remain perturbatively small. It also requires interatomic correlations, simultaneous interactions, and mode mixing to be negligible on the coarse-graining scale. The dependence on the chosen mode normalization enters the individual rates, whereas a common normalization cancels from their quotient.

Write \(p_{n,\nu}(t)\) for the probability of occupying photon number \(n\in\mathbb N_0\) in this mode, with \(p_{-1,\nu}=0\). The bosonic factors \(n+1\) for photon creation and \(n\) for absorption give the birth-death equation
\begin{equation} \dot p_{n,\nu}=\Gamma_{\mathrm e,\nu}[n p_{n-1,\nu}-(n+1)p_{n,\nu}]+\Gamma_{\mathrm a,\nu}[(n+1)p_{n+1,\nu}-n p_{n,\nu}]. \label{6.2} \end{equation}
For clarity, introduce the net probability current from level \(n\) to level \(n+1\), with boundary convention \(J_{-1,\nu}=0\),
\begin{align} J_{n,\nu}&=(n+1)[\Gamma_{\mathrm e,\nu}p_{n,\nu}-\Gamma_{\mathrm a,\nu}p_{n+1,\nu}],\nn \dot p_{n,\nu}&=J_{n-1,\nu}-J_{n,\nu}. \label{6.3} \end{align}
Summing Eq. \eqref{6.2} over \(n\) conserves total probability when the boundary currents vanish at large photon number. Define the mean occupation by \(N_\nu=\sum_{n=0}^{\infty}n p_{n,\nu}\). Multiplying the equation by \(n\) and summing gives the closed evolution equation
\begin{equation} \dot N_\nu=\Gamma_{\mathrm e,\nu}(N_\nu+1)-\Gamma_{\mathrm a,\nu}N_\nu=\Gamma_{\mathrm e,\nu}-(\Gamma_{\mathrm a,\nu}-\Gamma_{\mathrm e,\nu})N_\nu. \label{6.4} \end{equation}
Assume \(0<q_\nu<1\) and constant rates for the moment. For \(n\geq0\), stationarity, together with the vanishing current at the photon-number boundary, requires \(J_{n,\nu}=0\). Hence
\begin{align} p_{n+1,\nu}^{\mathrm{ss}}&=q_\nu p_{n,\nu}^{\mathrm{ss}},\nn p_{n,\nu}^{\mathrm{ss}}&=(1-q_\nu)q_\nu^n. \label{6.5} \end{align}
Normalization follows from the convergent geometric series. The relaxation rate is \(\gamma_\nu=\Gamma_{\mathrm a,\nu}-\Gamma_{\mathrm e,\nu}>0\), and the stationary mean photon number is
\begin{equation} N_\nu^{\mathrm{ss}}=\frac{q_\nu}{1-q_\nu}. \label{6.6} \end{equation}
At fixed rates, Eq. \eqref{6.4} integrates to
\begin{equation} N_\nu(t)=N_\nu^{\mathrm{ss}}+[N_\nu(0)-N_\nu^{\mathrm{ss}}]e^{-\gamma_\nu t}. \label{6.7} \end{equation}
The positive relaxation rate establishes stability of the mean occupation. The Markov birth-death process also admits the normalizable geometric stationary state of Eq. \eqref{6.5}, provided the initial probability distribution and its relevant moments remain well defined. A number-changing radiative flux can occur during approach to stationarity, but the mean occupation ceases to change at the stationary point of this closed gain-loss model.

In the high-gap limit established in Eq. \eqref{4.15}, the spectral quotient becomes \(q_\nu^{\mathrm H}=e^{-2\pi\nu/\kappa}\). The stationary mean is then
\begin{equation} N_\nu^{\mathrm H}=\frac{1}{e^{2\pi\nu/\kappa}-1}. \label{6.8} \end{equation}
The Planck occupation law thus follows from the independent-atom dynamics and the thermal transition ratio, rather than from the gamma-function factor alone \cite{Scully:2017utk,Azizi:2021qcu}. At finite gap or interaction-envelope width, Eq. \eqref{3.19} instead gives a mode-dependent geometric stationary state without requiring a common temperature. Replacing that quotient by the quadratic-jet approximation of Eq. \eqref{5.14} is consistent only where its probability corrections and the corresponding worldline remainder are small.

\subsection{Radiative entropy and nonequilibrium balance}

For a diagonal single-mode state, the von Neumann entropy equals the Shannon entropy of its photon-number probabilities,
\begin{equation} S_\nu(t)=-k_B\sum_{n=0}^{\infty}p_{n,\nu}(t)\ln p_{n,\nu}(t). \label{6.9} \end{equation}
We retain \(k_B\) in entropy formulas to make the physical normalization transparent, although it equals unity in our geometric-unit calculations. Substituting the geometric stationary probabilities gives
\begin{equation} S_\nu^{\mathrm{ss}}=k_B[(N_\nu^{\mathrm{ss}}+1)\ln(N_\nu^{\mathrm{ss}}+1)-N_\nu^{\mathrm{ss}}\ln N_\nu^{\mathrm{ss}}]. \label{6.10} \end{equation}
Define \(\Lambda_\nu=-\ln q_\nu>0\). Differentiating the geometric entropy and evaluating its slope at the stationary mean occupation gives
\begin{equation} \left.\frac{dS_\nu^{\mathrm{geom}}}{dN_\nu}\right|_{N_\nu=N_\nu^{\mathrm{ss}}}=k_B\ln\!\left(\frac{N_\nu^{\mathrm{ss}}+1}{N_\nu^{\mathrm{ss}}}\right)=k_B\Lambda_\nu. \label{6.11} \end{equation}
The last equality in Eq. \eqref{6.11} uses the stationary relation between mean occupation and the microscopic rate quotient. A transient geometric distribution with a different mean does not generally have that same entropy slope. More generally, a transient photon distribution need not remain geometric, even if its mean obeys Eq. \eqref{6.4}. We therefore derive the entropy rate directly from the master equation rather than assuming that the stationary functional form remains exact during evolution.

Suppose the probabilities are strictly positive and decrease sufficiently rapidly with \(n\) to justify the following summations and boundary limits. Probability conservation and Eq. \eqref{6.3} give
\begin{equation} \dot S_\nu=-k_B\sum_{n=0}^{\infty}\dot p_{n,\nu}\ln p_{n,\nu}=k_B\sum_{n=0}^{\infty}J_{n,\nu}\ln\!\left(\frac{p_{n,\nu}}{p_{n+1,\nu}}\right). \label{6.12} \end{equation}
Insert the identity \(\ln(p_{n,\nu}/p_{n+1,\nu})=\Lambda_\nu+\ln[\Gamma_{\mathrm e,\nu}p_{n,\nu}/(\Gamma_{\mathrm a,\nu}p_{n+1,\nu})]\). Since \(\sum_nJ_{n,\nu}=\dot N_\nu\), we obtain the exact decomposition
\begin{align} \dot S_\nu&=k_B\Lambda_\nu\dot N_\nu+\Pi_\nu,\nn \Pi_\nu&=k_B\sum_{n=0}^{\infty}J_{n,\nu}\ln\!\left(\frac{\Gamma_{\mathrm e,\nu}p_{n,\nu}}{\Gamma_{\mathrm a,\nu}p_{n+1,\nu}}\right). \label{6.13} \end{align}
Each term of \(\Pi_\nu\) is nonnegative, because \(J_{n,\nu}\) has the sign of \(\Gamma_{\mathrm e,\nu}p_{n,\nu}-\Gamma_{\mathrm a,\nu}p_{n+1,\nu}\) and the logarithm has the same sign. Therefore,
\begin{equation} \Pi_\nu\geq0. \label{6.14} \end{equation}
For the geometric stationary distribution, detailed balance gives \(\Pi_\nu=0\). The quantity \(\Pi_\nu\) is the irreversible entropy-production term of the birth-death dynamics \cite{Schnakenberg:1976zz,Spohn:1978pfz}. Its presence shows why the relation \(\dot S_\nu=k_B\Lambda_\nu\dot N_\nu\) is not generally exact for a transient state. It also shows that exact stationarity entails \(\dot N_\nu=\dot S_\nu=0\), even though the stationary entropy itself is nonzero.

A quantitative near-stationary estimate follows without assuming an exactly geometric transient state. Write \(\pi_{n,\nu}=(1-q_\nu)q_\nu^n\) for the instantaneous stationary reference distribution, and suppose the perturbations satisfy \(\sum_{n=0}^{\infty}\pi_{n,\nu}\epsilon_{n,\nu}=0\) and \(|\epsilon_{n,\nu}|\leq\epsilon<1\):
\begin{equation} p_{n,\nu}=\pi_{n,\nu}(1+\epsilon_{n,\nu}). \label{6.15} \end{equation}
Detailed balance for \(\pi_{n,\nu}\) implies \(J_{n,\nu}=(n+1)\Gamma_{\mathrm e,\nu}\pi_{n,\nu}(\epsilon_{n,\nu}-\epsilon_{n+1,\nu})\). Applying the mean-value bound to the logarithm in Eq. \eqref{6.13}, and then using \(\sum_n(n+1)\pi_{n,\nu}=1/(1-q_\nu)\), gives
\begin{equation} 0\leq\Pi_\nu\leq\frac{4k_B\Gamma_{\mathrm e,\nu}\epsilon^2}{(1-\epsilon)(1-q_\nu)}. \label{6.16} \end{equation}
Thus the omitted entropy-production term is second order in the relative departures from instantaneous stationarity, under the stated uniform bound. By contrast, \(\dot N_\nu\) can be first order in those departures. The same instantaneous identity remains valid when the transition rates vary slowly in time, although approaching the moving stationary distribution then requires the appropriate separation between relaxation and driving timescales.

For statistically independent modes with a product diagonal state and convergent entropy sums, the total radiative entropy and its rate are additive. Define \(\Pi_{\mathrm{tot}}=\sum_\nu\Pi_\nu\geq0\). The result is
\begin{equation} \dot S_p=k_B\sum_\nu\Lambda_\nu\dot N_\nu+\Pi_{\mathrm{tot}}. \label{6.17} \end{equation}
This identity is an exact property of the specified Markov model \cite{Schnakenberg:1976zz,Spohn:1978pfz}. Its near-stationary version, with \(\Pi_{\mathrm{tot}}\) neglected under Eq. \eqref{6.16}, is the entropy-flux approximation used in the HBAR area argument \cite{Scully:2017utk,Azizi:2021yto}.

\subsection{Radiative area budget and finite-jet corrections}

The frequency \(\nu\) denotes the Killing energy per photon in geometric units. For the selected radiative subsystem, define the mode-energy change and its rate by
\begin{align} U_p&=\sum_\nu\nu N_\nu,\nn \dot U_p&=\sum_\nu\nu\dot N_\nu. \label{6.18} \end{align}
The sum is assumed finite or supplied with a finite physical mode range. Identifying \(\dot U_p\) with energy carried into an outgoing radiation sector requires that the counted change represent outward photons rather than merely redistribution or accumulation inside a cavity. In particular, a stationary trapped mode with \(\dot N_\nu=0\) does not acquire a nonzero escaping luminosity merely from the gain-loss equation. Additional photon leakage or scattering mechanisms would have to be included in the dynamics before such a luminosity could be inferred.

The Schwarzschild horizon area and its derivative with respect to the mass are
\begin{align} A&=16\pi M^2,\nn \frac{dA}{dM}&=32\pi M,\nn \kappa&=\frac{1}{4M}. \label{6.19} \end{align}
Following the radiative-component convention of the original HBAR entropy argument \cite{Bekenstein:1973ur,Scully:2017utk,Azizi:2021yto}, define a positive-oriented area-equivalent quantity from the radiative energy rate,
\begin{equation} \dot A_p^{(+)}=32\pi M\dot U_p=\frac{8\pi}{\kappa}\sum_\nu\nu\dot N_\nu. \label{6.20} \end{equation}
The superscript \((+)\) refers to the positive radiative orientation of the area budget. Its rate is positive when the identified net radiative energy rate is positive. The definition does not assert that the actual black-hole horizon grows when photons escape. Its role is to express the radiation energy scale through the area derivative of the Schwarzschild family.

By Eq. \eqref{6.20}, the thermal part of the exact entropy balance can be rewritten as
\begin{equation} \frac{k_B}{4}\dot A_p^{(+)}=\frac{2\pi k_B}{\kappa}\sum_\nu\nu\dot N_\nu. \label{6.21} \end{equation}
Combining Eqs. \eqref{6.17} and \eqref{6.21} yields the central conditional relation
\begin{equation} \dot S_p=\frac{k_B}{4}\dot A_p^{(+)}+k_B\sum_\nu\left[\Lambda_\nu-\frac{2\pi\nu}{\kappa}\right]\dot N_\nu+\Pi_{\mathrm{tot}}. \label{6.22} \end{equation}
The identity separates three quantities with different meanings. The first is the Hawking-coefficient area equivalent of the radiative energy rate. The second is the departure of microscopic detailed balance from the ideal thermal exponent. The third is the irreversible entropy production associated with the radiation distribution not being exactly geometric. None of the terms requires a newly assigned proper acceleration for the freely falling atoms.

In the regime of Eq. \eqref{6.15}, let a common \(\epsilon\) bound hold over the retained modes and suppose the indicated sum converges. The exact decomposition then gives the near-stationary relation and an explicit error estimate,
\begin{align} \dot S_p&=\frac{k_B}{4}\dot A_p^{(+)}+k_B\sum_\nu\left[\Lambda_\nu-\frac{2\pi\nu}{\kappa}\right]\dot N_\nu+\mathcal E_{\mathrm{ns}},\nn 0&\leq\mathcal E_{\mathrm{ns}}\leq\frac{4k_B\epsilon^2}{1-\epsilon}\sum_\nu\frac{\Gamma_{\mathrm e,\nu}}{1-q_\nu}. \label{6.23} \end{align}
In the exact entropy balance, the remainder \(\mathcal E_{\mathrm{ns}}\) is \(\Pi_{\mathrm{tot}}\). The bound in Eq. \eqref{6.23} is absolute. A leading relative approximation also requires \(\Pi_{\mathrm{tot}}\ll\left|k_B\sum_\nu\Lambda_\nu\dot N_\nu\right|\) whenever the quantity on the right is nonzero. Near stationarity alone does not ensure this comparison, because the weighted occupation-change sum can vanish or become arbitrarily small while entropy production remains positive. If that sum vanishes, Eq. \eqref{6.23} supplies the appropriate absolute statement and no relative accuracy follows.
In the ideal high-gap thermal limit, \(q_\nu=e^{-2\pi\nu/\kappa}\) and hence \(\Lambda_\nu=2\pi\nu/\kappa\), so the spectral correction in Eq. \eqref{6.22} vanishes. When the relative condition stated above also holds and \(\Pi_{\mathrm{tot}}\) is negligible compared with the leading entropy-transport term, the near-stationary relation reduces to
\begin{equation} \dot S_p\simeq\frac{k_B}{4}\dot A_p^{(+)}. \label{6.24} \end{equation}
The relation reproduces the HBAR entropy-area coefficient within the original near-steady interpretation (\cite{Scully:2017utk}). It does not imply a nonzero entropy derivative at exact stationarity, where every current vanishes, or establish an area identity for an arbitrary far-from-equilibrium state.

The finite-envelope and worldline-jet calculations specify the spectral part of the correction without imposing a new equilibrium temperature. For the linear jet, \(\Lambda_\nu=\Lambda_{\mathrm{lin}}(\nu)\) is given exactly by Eq. \eqref{5.20}. At first order in the quadratic spectral correction of Eq. \eqref{5.18}, the additional term is
\begin{align} \Delta\dot S_{p,\mathrm{jet}}^{[2]}&=-k_B\sum_\nu\Delta_2(\nu)\dot N_\nu
\nn &+\mathcal{O}\!\left(k_B\sum_\nu[|\delta_{\mathrm e,\nu}^{(2)}|^2+|\delta_{\mathrm a,\nu}^{(2)}|^2]|\dot N_\nu|\right). \label{6.25} \end{align}
The expression isolates the explicit effect of the quadratic-jet correction on the logarithmic spectral ratio. A calculation of the complete change in entropy rate must also account for the induced changes in \(\Gamma_{\mathrm e,\nu}\), \(\Gamma_{\mathrm a,\nu}\), \(\dot N_\nu\), and the state itself. Moreover, Eq. \eqref{6.25} inherits the omitted higher-jet and remote-interaction uncertainties identified in Section 5. The spectral contribution is therefore conditional and does not constitute an exact correction to black-hole thermodynamics.

Restore physical units by treating \(\nu\) as an angular frequency measured relative to Killing time at infinity. If \(\dot E_p=\hbar\sum_\nu\nu\dot N_\nu\) is identified with the radiative energy flux, the area-equivalent rate becomes
\begin{align} \dot A_p^{(+)}&=\frac{32\pi G^2M}{c^6}\dot E_p,\nn \dot E_p&=\hbar\sum_\nu\nu\dot N_\nu. \label{6.26} \end{align}
The ideal HBAR relation consequently reads
\begin{equation} \dot S_p\simeq\frac{k_Bc^3}{4\hbar G}\dot A_p^{(+)}. \label{6.27} \end{equation}
For comparison, the actual signed horizon-area rate is \(\dot A=32\pi G^2M\dot M/c^4\), where \(\dot M\) is set by the net mass-energy exchange including the atomic influx and all outgoing and ingoing channels. The positive-oriented radiative area equivalent in Eq. \eqref{6.26} is not, by itself, that signed rate. The entropy \(S_p\) belongs to the radiation subsystem and must not be identified with the intrinsic Bekenstein-Hawking entropy.

The analytical hierarchy of this section consequently links the affine transition quotient to a geometric photon-number distribution, an exact Markov entropy balance, and a conditional radiative area equivalent. Thermal detailed balance recovers the HBAR area coefficient, while finite interaction envelopes and successive worldline jets add explicit spectral terms. The description remains limited to the independent-mode weak-coupling approximation and the near-horizon local response established earlier. A cavity with physical boundaries, outgoing photon escape, global mode normalization, or coupled photon channels requires a corresponding extension of the master equation before an observed radiative flux can be inferred.

\section{Worldline-Jet Spectra in the Simpson-Visser Black-Bounce Geometry}
\label{sec7}

The preceding analysis associates the thermal leading term with the logarithmic outgoing phase and associates the regular worldline coefficients with deviations from that term. We now apply the same proper-time jet and Mellin-Laplace moment procedure to the Simpson-Visser geometry \cite{Simpson:2018tsi}. A nonextremal HBAR calculation in this spacetime has already recovered the surface-gravity temperature and examined the radiative entropy \cite{Ovgun:2026plj}. Our purpose is different. We retain the arbitrary infall energy and calculate how the first two regular coefficients enter finite-envelope emission and absorption. The critical null throat supplies a separate test of the domain in which the affine representation exists. We keep the same local outgoing scalar-mode approximation and weak atom-field interaction used in Sections 3-5.

\subsection{Nondegenerate horizon and the proper-time jet}

Write \(R=2M>0\) and denote the nonnegative bounce parameter by \(a\). The static metric and lapse are
\begin{align} d\ell^2&=-F(r)dt^2+\frac{dr^2}{F(r)}+(r^2+a^2)d\Omega_2^2,\nn F(r)&=1-\frac{R}{\sqrt{r^2+a^2}}. \label{7.1} \end{align}
Introduce the dimensionless parameter \(h=a^2/R^2\). For \(0\leq a<R\), equivalently \(0\leq h<1\), the positive exterior horizon is nondegenerate. The horizon coordinate and its surface gravity are
\begin{align} r_h&=R\sqrt{1-h},\nn \kappa_a&=\frac{F'(r_h)}{2}=\frac{\sqrt{1-h}}{2R}. \label{7.2} \end{align}
The areal radius of the horizon equals \(R\), whereas its coordinate location and surface gravity vary with \(a\). The limit \(a=0\) reduces the exterior positive-\(r\) branch to Schwarzschild. For fixed \(E>0\), the infalling timelike geodesic satisfies the same radial identity as Eq. \eqref{2.3}, with \(F\) replacing \(f\). Its remaining proper time is measured from crossing \(r_h\), and its outgoing retarded time is defined by \(du=dt-dr/F(r)\). Consequently, $r(0)=r_h$ and
\begin{align} \frac{dr}{ds}&=\sqrt{E^2-F(r)},\nn \frac{du}{ds}&=-\frac{1}{E-\sqrt{E^2-F(r)}}. \label{7.3} \end{align}
The regularity of the radial square root near a simple horizon permits a local Taylor expansion. Introduce \(F_j=F^{(j)}(r_h)\) for \(j=1,2,3\). Repeated differentiation of \(r''=-F'(r)/2\) yields
\begin{align} r(s)&=r_h+Es-\frac{F_1}{4}s^2-\frac{EF_2}{12}s^3
\nn &+\left(\frac{F_1F_2}{96}-\frac{E^2F_3}{48}\right)s^4+\mathcal{O}_{E,h}(s^5/R^4). \label{7.4} \end{align}
Substitution into the rationalized outgoing-time identity of Eq. \eqref{7.3} gives the universal simple-horizon jet
\begin{equation} u(s)=u_0-\kappa_a^{-1}\ln(s/s_0)+c_1s+c_2s^2+\mathcal{O}_{E,h}(s^3/R^2). \label{7.5} \end{equation}
Here the remainder constants depend on the fixed geometry and geodesic energy. The regular coefficients are determined without integrating an exact radial trajectory,
\begin{align} c_1&=\frac{EF_2}{F_1^2},\nn c_2&=\frac{E^2(2F_1F_3-3F_2^2)-F_1^2F_2}{12F_1^3}. \label{7.6} \end{align}
For the Simpson-Visser lapse, evaluation at the positive horizon gives
\begin{align} F_1&=\frac{\sqrt{1-h}}{R},\nn F_2&=\frac{3h-2}{R^2},\nn F_3&=\frac{3\sqrt{1-h}(2-5h)}{R^3}. \label{7.7} \end{align}
The two coefficients therefore take the explicit forms
\begin{align} c_1(E,h)&=\frac{E(3h-2)}{1-h},\nn c_2(E,h)&=\frac{(1-h)(2-3h)-3E^2h(2-h)}{12R(1-h)^{3/2}}. \label{7.8} \end{align}
At \(h=0\), these reduce to \(c_1=-2E\) and \(c_2=1/(6R)\), reproducing Eq. \eqref{2.13} with \(R=r_s\). More generally, the logarithmic coefficient depends only on \(\kappa_a\), while the regular coefficients depend independently on \(E\) and on higher derivatives of \(F\). The choices \(0<E<1\), \(E=1\), and \(E>1\) correspond, respectively, to suitable finite-radius release from rest, release from rest at infinity, and inward motion from infinity with nonzero asymptotic speed. All three share the same simple-horizon logarithm.

\subsection{Emission and absorption from the same Mellin moments}

Retain the one-sided envelope \(e^{-\alpha s}\) with \(\alpha>0\), the atomic gap \(\omega>0\), and an outgoing Killing frequency \(\nu>0\). Define the shifted frequencies and Laplace variables from the linear coefficient in Eq. \eqref{7.8},
\begin{align} \sigma_a&=\frac{\nu}{\kappa_a},\nn \Omega_{\mathrm e}^{a}&=\omega-\nu c_1,\nn \Omega_{\mathrm a}^{a}&=\omega+\nu c_1,\nn z_j^{a}&=\alpha+i\Omega_j^{a}. \label{7.9} \end{align}
The imaginary proper-time powers in Eqs. \eqref{3.8} and \eqref{3.9} remain unchanged in form, with \(\sigma\) replaced by \(\sigma_a\). The Mellin-Laplace identity of Eq. \eqref{3.12} therefore gives the linear-jet amplitudes
\begin{align} \mathcal A_{\mathrm e}^{a,\mathrm{lin}}&=-ig\,e^{i\omega\tau_H+i\nu u_0}s_0^{i\sigma_a}\Gamma(1-i\sigma_a)(z_{\mathrm e}^{a})^{-1+i\sigma_a},\nn \mathcal A_{\mathrm a}^{a,\mathrm{lin}}&=-ig\,e^{i\omega\tau_H-i\nu u_0}s_0^{-i\sigma_a}\Gamma(1+i\sigma_a)(z_{\mathrm a}^{a})^{-1-i\sigma_a}. \label{7.10} \end{align}
Let \(\theta_j^{a}=\arctan(\Omega_j^{a}/\alpha)\). The same complex-power calculation as in Section 3 then yields
\begin{equation} q_a^{\mathrm{lin}}(\nu)=\frac{\alpha^2+(\Omega_{\mathrm a}^{a})^2}{\alpha^2+(\Omega_{\mathrm e}^{a})^2}\exp[-2\sigma_a(\theta_{\mathrm e}^{a}+\theta_{\mathrm a}^{a})]. \label{7.11} \end{equation}
The quotient is exact within the linear-jet and constant-amplitude outgoing-mode model. For both shifted frequencies positive, its high-gap expansion follows by setting \(\eta_a=-\nu c_1/\omega\), \(\delta=\alpha/\omega\), and \(\varepsilon=\max\{|\eta_a|,\delta\}\). At fixed \(\sigma_a\) and $\varepsilon\to0$,
\begin{equation} \ln q_a^{\mathrm{lin}}=-2\pi\sigma_a+\frac{4\nu c_1}{\omega}+\frac{4\sigma_a\alpha}{\omega}+\mathcal{O}_{\sigma_a}(\varepsilon^3). \label{7.12} \end{equation}
Thus the familiar surface-gravity exponent is recovered, while the first finite-gap term can change sign when \(c_1\) does. Neither its sign nor its size is determined by surface gravity alone. Equation \eqref{7.12} assumes the spectral frequencies and the energy are held in the specified scale-separated regime. It does not describe a uniform approach to the critical null throat.

The next worldline coefficient enters through precisely the degree-two moment of Eq. \eqref{5.5}. Expanding the additional factor \(\exp[\pm i\nu c_2s^2]\) gives
\begin{align} \delta_{\mathrm e}^{a,(2)}&=i\nu c_2\frac{(1-i\sigma_a)(2-i\sigma_a)}{(z_{\mathrm e}^{a})^2},\nn \delta_{\mathrm a}^{a,(2)}&=-i\nu c_2\frac{(1+i\sigma_a)(2+i\sigma_a)}{(z_{\mathrm a}^{a})^2}. \label{7.13} \end{align}
The corrected emission-to-absorption quotient is consequently
\begin{align} \widetilde q_a^{[2]}&=q_a^{\mathrm{lin}}[1+\Delta_{2,a}(\nu)+\mathcal{O}(|\delta_{\mathrm e}^{a,(2)}|^2+|\delta_{\mathrm a}^{a,(2)}|^2)],\nn \Delta_{2,a}&=2\operatorname{Re}(\delta_{\mathrm e}^{a,(2)}-\delta_{\mathrm a}^{a,(2)}). \label{7.14} \end{align}
The same Mellin family has thus supplied both the finite-envelope quotient and the first nonlinear geometrical correction. In particular, replacing \(\kappa\) by \(\kappa_a\) in a Schwarzschild Planck factor would not produce Eq. \eqref{7.13}. The coefficient \(c_2(E,h)\) contains geometric information absent from surface gravity and from the linear coefficient alone.

The local interpretation requires a separate error estimate. Let a smooth function \(h_L(s)\) equal unity for \(0\leq s\leq L\), vanish for \(s\geq2L\), and lie between zero and unity. For fixed \(h<1\), choose \(L\) so that the geodesic jet is valid on \(0\leq s\leq2L\) and \(F(r(s))\ll E^2\). Suppose \(|u(s)-u_{[2]}(s)|\leq C_{E,h}s^3\) there. If \(\mathcal A_j^{a,h_L}\) is the amplitude using the exact exterior trajectory and the localized envelope \(h_L(s)e^{-\alpha s}\), then, for either channel \(j=\mathrm e,\mathrm a\), the phase remainder and the tail integral give
\begin{equation} \left|\mathcal A_j^{a,h_L}-\widetilde{\mathcal A}_j^{a,[2]}\right|\leq g\left[\frac{6C_{E,h}\nu}{\alpha^4}+\frac{12\nu^2c_2^2}{\alpha^5}+\frac{e^{-\alpha L}}{\alpha}\right]. \label{7.15} \end{equation}
The estimate is an absolute one and is useful for a probability quotient only if it is small relative to both leading amplitude magnitudes. It excludes changes in the radial mode normalization, scattering from the effective potential, and physical cavity boundaries. A sufficiently localized envelope requires the existence of an interval with \(\alpha L\gg1\), while the high-gap use of Eq. \eqref{7.12} also requires \(\omega\gg\alpha\).

\subsection{A vanishing linear coefficient and its energy dependence}

The expression for \(c_1\) has an exact zero within the nondegenerate black-bounce branch. Its parameter value and surface gravity are
\begin{align} a_\star&=R\sqrt{\frac{2}{3}},\nn h_\star&=\frac{2}{3},\nn \kappa_\star&=\frac{1}{2\sqrt{3}R}. \label{7.16} \end{align}
At this point the linear regular phase disappears for every \(E>0\), but the quadratic coefficient remains nonzero and depends on the square of that energy,
\begin{align} c_1(E,h_\star)&=0,\nn c_2(E,h_\star)&=-\frac{2E^2}{\sqrt{3}R}. \label{7.17} \end{align}
At this parameter value, define \(\sigma_\star=\nu/\kappa_\star\). The linear-jet quotient simplifies without taking the high-gap limit,
\begin{equation} q_\star^{\mathrm{lin}}(\nu)=\exp\!\left[-4\sigma_\star\arctan\!\left(\frac{\omega}{\alpha}\right)\right]. \label{7.18} \end{equation}
Within the linear-jet model, Eq. \eqref{7.18} is exactly exponential in the Killing frequency, with \(q_\star^{\mathrm{lin}}(\nu)=e^{-\beta_{\mathrm{model}}\nu}\) and \(\beta_{\mathrm{model}}=4\kappa_\star^{-1}\arctan(\omega/\alpha)\). At finite \(\alpha>0\), this parameter differs from the Hawking inverse temperature \(\beta_H=2\pi/\kappa_\star\). The Abel limit \(\alpha\to0^+\) recovers \(\beta_H\). The quadratic Mellin moments identify the first explicit energy dependence in the constant-amplitude local model.
\begin{equation} \Delta_{2,\star}(\nu)=-\frac{16E^2\alpha\omega\nu(2-\sigma_\star^2)}{\sqrt{3}R(\alpha^2+\omega^2)^2}. \label{7.19} \end{equation}
The frequency dependence of \(\Delta_{2,\star}(\nu)\) introduces a departure from the exact exponential form at the order retained. For \(E=1/2\), \(E=1\), and \(E=2\), the coefficient \(c_2\) has the relative magnitudes \(1\), \(4\), and \(16\). These cases show how the same geometry enters the quadratic transition moments differently for different infalling atoms. The coefficient in Eq. \eqref{7.19} is not the full exact-worldline correction. The cubic geodesic jet and the varying radial wave amplitude can contribute separately, and the quadratic probability expansion is valid only when its relative corrections are small. The disappearance of \(c_1\) therefore does not eliminate all energy dependence of a globally normalized field response.

\subsection{Critical null throat and the limit of affine scaling}

The nondegenerate formulas cannot simply be evaluated at \(a=R\). At the critical value, the horizons coalesce at \(r=0\), \(\kappa_a\) vanishes, and the leading lapse is quadratic rather than linear \cite{Simpson:2018tsi}. On the positive exterior branch,
\begin{equation} F(r)=\frac{r^2}{2R^2}-\frac{3r^4}{8R^4}+\mathcal{O}(r^6/R^6). \label{7.20} \end{equation}
The same radial equation as Eq. \eqref{7.3} remains regular for every fixed \(E>0\). Expanding it through the first nonlinear contribution and then integrating the outgoing-time derivative gives
\begin{align} r(s)&=Es-\frac{E}{12R^2}s^3+\mathcal{O}_E(s^5/R^4),\nn u(s)&=u_0+\frac{4R^2}{Es}-\left(3E+\frac{1}{6E}\right)s+\mathcal{O}_E(s^3/R^2). \label{7.21} \end{align}
The essential distinction is the inverse-power phase in place of the logarithmic one. Define \(C=4R^2\nu/E>0\), \(\gamma_E=3E+1/(6E)\), and the shifted Laplace variables associated with the retained regular term,
\begin{equation} z_{\mathrm e}^{\mathrm{crit}}=\alpha+i(\omega+\nu\gamma_E),\qquad z_{\mathrm a}^{\mathrm{crit}}=\alpha+i(\omega-\nu\gamma_E). \label{7.22} \end{equation}
The leading critical-jet amplitudes are no longer Mellin transforms. Nevertheless, applying the same proper-time phase expansion before evaluating the integral leads to a closed inverse-phase Laplace expression,
\begin{equation} \mathcal I_{\mathrm e}^{\mathrm{crit}}=\int_0^\infty ds\,e^{-z_{\mathrm e}^{\mathrm{crit}}s+iC/s}=2\sqrt{\frac{-iC}{z_{\mathrm e}^{\mathrm{crit}}}}K_1\!\left(2\sqrt{-iCz_{\mathrm e}^{\mathrm{crit}}}\right). \label{7.23} \end{equation}
The opposite field phase gives
\begin{equation} \mathcal I_{\mathrm a}^{\mathrm{crit}}=\int_0^\infty ds\,e^{-z_{\mathrm a}^{\mathrm{crit}}s-iC/s}=2\sqrt{\frac{iC}{z_{\mathrm a}^{\mathrm{crit}}}}K_1\!\left(2\sqrt{iCz_{\mathrm a}^{\mathrm{crit}}}\right). \label{7.24} \end{equation}
Here $K_1$ denotes the modified Bessel function of the second kind. Both formulas follow from the modified-Bessel integral representation by analytic continuation \cite{DLMF}, with \(\alpha>0\) and the principal square-root branch. The corresponding elementary probability ratio in this leading critical-jet model is
\begin{equation} q_{\mathrm{crit}}^{[1]}=\frac{\left|\mathcal I_{\mathrm e}^{\mathrm{crit}}\right|^2}{\left|\mathcal I_{\mathrm a}^{\mathrm{crit}}\right|^2}. \label{7.25} \end{equation}
Its functional dependence is not the gamma-function quotient of Eq. \eqref{7.11}, and it does not furnish a thermal detailed-balance law from the simple-horizon argument. The integral extension to the entire positive half-line is a formal local model at this critical geometry. A local comparison is possible using the same smooth envelope \(h_L\) defined above. For a sufficiently small \(L/R\), suppose the first omitted critical-jet phase satisfies \(|u(s)-u_{\mathrm{crit},[1]}(s)|\leq D_Es^3/R^2\) for \(0\leq s\leq2L\), where \(u_{\mathrm{crit},[1]}\) denotes the two displayed terms of Eq. \eqref{7.21} following \(u_0\). Then, for either channel \(j=\mathrm e,\mathrm a\), each exact-worldline amplitude with the additional cutoff satisfies
\begin{equation} \left|\mathcal A_j^{\mathrm{crit},h_L}-\mathcal A_j^{\mathrm{crit},[1]}\right|\leq g\left[\frac{6D_E\nu}{R^2\alpha^4}+\frac{e^{-\alpha L}}{\alpha}\right]. \label{7.26} \end{equation}
Here \(\mathcal A_j^{\mathrm{crit},[1]}\) includes the constant reference phases and the corresponding integral in Eqs. \eqref{7.23} and \eqref{7.24}. The bound is absolute and is not, without an additional lower bound on the amplitudes, a relative error estimate for \(q_{\mathrm{crit}}^{[1]}\). The remaining phase terms and the field state at a null throat require independent treatment before interpreting the ratio as physical radiation.

The failure of a direct limiting substitution has an explicit geometric cause. As \(h\to1^-\), the nondegenerate horizon coordinate \(r_h=R\sqrt{1-h}\) shrinks to zero, while \(c_1\) and \(c_2\) in Eq. \eqref{7.8} become singular and \(\kappa_a\to0\). The interval where the simple-horizon logarithm dominates the quadratic lapse shrinks with \(r_h/E\). Hence the limits \(s\to0^+\) at fixed \(h<1\) and \(h\to1^-\) at fixed proper time cannot be interchanged in Eqs. \eqref{7.5}-\eqref{7.14}. For \(a>R\), no Killing horizon exists and a horizon-crossing proper-time variable of the present kind is unavailable \cite{Simpson:2018tsi}.

The leading HBAR Planck factor and surface-gravity temperature, the affine translation-dilation argument, finite-gate departures from thermality, the leading entropy-area coefficient, and the nonextremal Simpson-Visser HBAR response are established results \cite{Scully:2017utk,Camblong:2020pme,Azizi:2021yto,Arzano:2025aeo,Arzano:2026vwz,Schlicht:2003iy,Louko:2006zv,Satz:2006kb,Louko:2007mu,Lobos:2026ahg,Ovgun:2026plj}. We use these results as the physical and mathematical starting point rather than claim them as new. The present contribution is a complementary local spectral treatment that carries successive proper-time coefficients of the infalling trajectory through a common Mellin-Laplace moment family. For Schwarzschild, the principal additions are the quadratic-jet correction to the transition quotient in Eq. \eqref{5.18}, the exact linear-jet low-frequency coefficients, the first-order quadratic-jet contributions, and the cubic worldline moment in Eqs. \eqref{5.22}--\eqref{5.27}. The coefficients \(\delta b_1^{(2)}\) and \(\delta b_3^{(2)}\) are partial quadratic-phase terms and do not give the full frequency coefficients of the exact trajectory. For Simpson-Visser, the corresponding additions are the general simple-horizon coefficients in Eq. \eqref{7.8}, the distinguished-parameter results in Eqs. \eqref{7.17}--\eqref{7.19}, and the critical-throat inverse-phase integrals with their local absolute error estimate in Eqs. \eqref{7.21}--\eqref{7.26}. These results concern the selected outgoing mode with frozen radial amplitude and do not establish a global radiation spectrum, a new thermal law, or a new horizon-entropy relation.

\section{Conclusions}\label{sec8}

We have developed a proper-time worldline-jet formulation of the outgoing-mode response of a freely falling two-level atom in the Schwarzschild exterior. The radial geodesic determines a logarithmic retarded-time phase whose leading coefficient is fixed by the surface gravity. Rewriting the resulting transition integrals in Fourier-Mellin form provides a common gamma-function representation for photon emission and absorption. Within a one-sided exponential interaction model, the resulting probability quotient retains the atomic gap, the conserved geodesic energy, and the envelope scale. In the localized high-gap regime, it approaches the established HBAR detailed-balance factor associated with the Schwarzschild Hawking temperature. The infalling atom remains geodesic throughout.

The spectral formulation also supports successive local corrections without changing the integral family. We have used the quadratic worldline coefficient to obtain explicit gamma-function moments and the leading change in the emission-to-absorption quotient. A separate small-frequency expansion of the linear-jet logarithmic ratio identifies both its initial slope and its cubic spectral coefficient. The latter is generically nonzero at finite gap and envelope width, so a single frequency-independent inverse temperature does not describe the full response of this model. Higher worldline coefficients and the finite extent of the interaction restrict the interpretation of these terms. In particular, the quadratic contribution to spectral curvature is not the complete curvature of the exact Schwarzschild response.

For a dilute beam of independently injected atoms, we have related the microscopic transition probabilities to a birth-death equation for the photon-number distribution. Its geometric stationary state becomes Planckian in the ideal HBAR limit. An exact entropy-balance identity separates the term associated with the transition-rate quotient from nonnegative irreversible entropy production. Under a near-stationary approximation and an explicitly defined radiative area equivalent, the entropy rate has the familiar HBAR area coefficient. The finite-jet spectral contribution is conditional on the same statistical and energy-flow assumptions. Neither the radiative area equivalent nor the radiation entropy is identical to the actual signed horizon-area change or to the intrinsic Bekenstein-Hawking entropy.

The Simpson-Visser example extends the same jet and Mellin-moment procedure beyond Schwarzschild. For nondegenerate horizons, the linear and quadratic retarded-time coefficients depend explicitly on the bounce parameter and the infall energy. At one nondegenerate parameter value the first regular phase coefficient vanishes for every positive geodesic energy, whereas the next coefficient remains quadratic in that energy. The finite-envelope transition ratio and its first quadratic correction make this distinction explicit. At the critical null throat the logarithmic phase is replaced by an inverse-proper-time term, and the leading transition integrals instead involve modified Bessel functions. This change identifies a boundary of the affine near-horizon representation, not a new thermal law or an exact global radiation spectrum.

The present analysis is restricted to a selected outgoing massless scalar mode with a frozen radial amplitude, a one-sided exponential envelope, fixed geodesic energy, and weak atom-field coupling. Extending the calculation to a globally normalized scattering basis requires the radial mode functions and the interference of outgoing and ingoing branches. Further work should treat smooth finite interaction profiles, realistic cavity boundaries, photon leakage, the coupled mode dynamics needed to infer an escaping flux, and the extension of the jet analysis to other nonextremal horizons. These studies would determine which local spectral corrections persist in an asymptotically defined radiation flux.

\acknowledgments
R. P. and A. \"O would like to acknowledge networking support of the COST Actions CA22113 - Fundamental challenges in theoretical physics (THEORY-CHALLENGES), CA23130 - Bridging high and low energies in search of quantum gravity (BridgeQG), CA23115 - Relativistic Quantum Information (RQI) , CA24139 - Superfluid Condensates in Astrophysics and Laboratory Experiments, and CA24149 - Extreme Plasma Network for Advanced Discovery (EXPAND), which are all funded by COST (European Cooperation in Science and Technology). R. P. and A. \"O. also thanks to EMU, TUBITAK, ULAKBIM (Turkiye) and SCOAP$^3$ (Switzerland) for their support.

\bibliography{main}

\end{document}